\documentclass[12pt]{article}
\usepackage{graphicx}
\usepackage[cp1251]{inputenc}
\usepackage{rotating}
\begin{document}
 \noindent {\footnotesize\it Astronomy Reports, 2021, Vol. 65, No 9, pp. 737--754}
 \newcommand{\dif}{\textrm{d}}

 \noindent
 \begin{tabular}{llllllllllllllllllllllllllllllllllllllllllllll}
 & & & & & & & & & & & & & & & & & & & & & & & & & & & & & & & & & & & & & &\\\hline\hline
 \end{tabular}

 \vskip 0.5cm
 \centerline{\bf\Large Comparison of the Orbital Properties of the Milky Way}
 \centerline{\bf\Large Globular Clusters from the Data of the Gaia DR2}
 \centerline{\bf\Large and EDR3 Catalogs}
 \bigskip
 \bigskip
  \centerline
 {
 A.T. Bajkova and V.V. Bobylev
 }
 \bigskip
 \centerline{\small \it
 Central (Pulkovo) Astronomical Observatory, Russian Academy of Sciences,}
 \centerline{\small \it Pulkovskoe shosse 65, St. Petersburg, 196140 Russia}
 \bigskip
 \bigskip
 \bigskip

 {
{\bf Abstract}---We provide new values of the orbital parameters
of 152 globular clusters, which are calculated using the new mean
proper motions obtained from the Gaia EDR3 catalog data. The
orbits were integrated 5 Gyr back in an axisymmetric
three-component potential with a spherical bulge, disk component,
and spherical dark halo in the Navarro–Frenk–White form, which we
refined using the rotation curve of objects with large
galactocentric distances up to 200 kpc. The obtained orbital
parameters were compared with the orbital parameters of the same
globular clusters calculated earlier in the same gravitational
potential using proper motions from the Gaia DR2 catalog data. The
objects whose orbits underwent significant changes were
identified.
 }

\medskip DOI: 10.1134/S1063772921090018

 \section{INTRODUCTION}
Globular star clusters (GCs) are among the most interesting
objects in our Galaxy. Studies of GCs provide understanding of the
birth and evolution of the Galaxy, since they are the oldest
stellar formations. Their age is almost equal to the age of the
Universe. Approximately 170 Milky Way GCs are currently known.
According to theoretical estimates, the number of GCs in the Milky
Way can be around 200.

One of the methods of GC studies is to examine their orbital
motion, which became possible thanks to high-accuracy measurements
of their spatial velocities and positions from the Gaia
spacecraft. The appearance of catalogs of mean proper motions
based on the second DR2 release in combination with other
astrometric data on GC radial velocities and positions made it
possible to study the orbital motion of nearly all GCs known to
date [1--3].

Among the astrometric data catalogs with proper motions from Gaia
DR2, we would like to specifically mention Vasiliev’s (2019)
catalog for 150 GCs [3], which allows the construction of the 6d
phase space required for the calculation of orbits. We used this
catalog to study the orbital properties of GCs. Thus, we developed
a new method for separating GCs by subsystems of the Galaxy:
bulge, thick disk, and halo [4]. The separation method is based on
the bimodality we discovered in the GC distribution over the
parameter $L_Z/ecc$, where  $L_Z$ is the $Z$ component of the
angular momentum and $ecc$ is the orbital eccentricity. Due to
this bimodality, GCs belonging to the disk, i.e., those that
formed in the Galaxy itself, can be easily distinguished from halo
GCs of extragalactic origin using the probabilistic method
described in detail in [4].

Massari et al. [5] give a classification of halo GCs that formed
outside the Milky Way and ended up in our Galaxy as a result of
accretion from dwarf galaxies (Sagittarius, Sausage, Sequoia [6]).
A catalog of 152 GC orbits and their orbital parameters is
presented by us in [7], as well as a modified classification by
the Galaxy subsystems based on the obtained orbital properties of
GCs. This classification is confirmed in this study as well.

The appearance of a new, more accurate version of the GC proper
motions catalog [8] based on the Gaia EDR3 measurement data leads
to a natural problem of refining the GC orbital parameters and
identifying GCs whose orbital motion underwent the greatest
changes. This is the scope of the present paper.

This study is structured as follows. The first section provides a
brief description and justification of the adopted model of the
gravitational potential in which the GC orbits are integrated; the
equations of motion and the formulas for calculating the orbital
parameters are also given. The second section describes the data
and compares the mean proper motions and their uncertainties
obtained from the data of the Gaia DR2 and EDR3 catalogs. The
third section is dedicated to the discussion of the study results;
a catalog of GC orbital parameters calculated from the Gaia EDR3
data is presented, and they are compared with the orbital
parameters obtained from the Gaia DR2 catalog and published in
[7]; the orbits for a number of GCs that have undergone the
greatest changes are given. The main conclusions are drawn in the
final section.

\section{METHODS}
\subsection{Axisymmetric Galactic Potential Model}
The axisymmetric gravitational potential of the Galaxy is
represented as a sum of three components: central spherical bulge
$\Phi_b(r(R,Z))$, disk $\Phi_d(r(R,Z))$, and a massive spherical
dark-matter halo $\Phi_h(r(R,Z))$ [9, 10]:
\begin{equation}
\begin{array}{lll}
  \Phi(R,Z)=\Phi_b(r(R,Z))+\Phi_d(r(R,Z))+\Phi_h(r(R,Z)).
 \label{pot}
 \end{array}
 \end{equation}
Here, we use a cylindrical coordinate system ($R,\psi,Z$) with the
origin at the center of the Galaxy. In a rectangular Cartesian
coordinate system $(X,Y,Z)$ with the origin at the center of the
Galaxy, the distance to the star (spherical radius) is
$r^2=X^2+Y^2+Z^2=R^2+Z^2$.

The bulge $\Phi_b(r(R,Z))$ and disk $\Phi_d(r(R,Z))$ potentials
are expressed in the form proposed by Miyamoto and Nagai [11], and
the halo component is represented according to Navarro et al.
[12]. The specific values of the parameters of this model,
provided that the potential is expressed in units of 100
km$^2$/s$^2$, distances in kpc, masses in units of the mass of the
Galaxy, $M_{gal}=2.325\times 10^7 M_\odot$, and the gravitational
constant $G=1$, are given in the paper by Bajkova and Bobylev [9],
where it is designated as model III.

It should be noted that the model of the galactic potential that
we have adopted and denoted here, same as in [7], by NFWBB, has
the parameters obtained by fitting them to the data on the
circular velocities of ionized hydrogen clouds HI, maser sources,
and various halo objects with large galactocentric distances $R$
up to $\sim200$~kpc from Bhattacharjee et al. [13] (see Fig. 1).
In addition, the fitting of the parameters involved the
constraints on the local dynamic density of matter $\rho_\odot=0.1
M_\odot$~pc$^{-3}$ and the force acting perpendicular to the plane
of the Galaxy $|K_{z=1.1}|/2\pi G=77M_\odot$~pc$^{-2}$ [14].

The rotation curve corresponding to the NFWBB model is shown in
Fig. 1. We plotted this curve using the values $R_\odot=8.3$~kpc
for the galactocentric distance of the Sun and $V_\odot=244$~km/s
for the rotation’s linear velocity of the local standard of rest
around the Galaxy center, as it is taken in [13]. According to
this model [9], the mass of the Galaxy
 $M_{G_{(R \leq 200~kpc)}}=0.75\pm0.19\times10^{12}M_\odot$. This value is in good
agreement with modern independent estimates. For example, the
lower estimate of the NFW halo mass obtained quite recently by
Koppelman and Helmi [15] from the data on the velocities of
runaway halo stars is
 $M_{G_{(R \leq 200~kpc)}}=0.67^{+0.30}_{-0.15}\times10^{12}M_\odot$. Figure 1, in
addition to the available data, also shows the circular velocities
of the thick disk GCs with orbital eccentricities <0.2 (blue
dots), which demonstrate good agreement with the data from maser
sources at interval of galactocentric distances $2<R<20$~kpc.

The model of the gravitational potential of the Milky Way NFWBB
appears to us the most realistic in comparison with other known
models, since it is supported by data at large galactocentric
distances, which is very important in integration of the orbits of
distant globular clusters and clusters with a large apocentric
distance; it also agrees well with modern estimates of the local
parameters and independent estimates of the mass of the Galaxy
[10], a thorough review of which was also recently given by Wang
et al. [16].

\begin{figure*}
\begin{center}
   \includegraphics[width=0.7\textwidth]{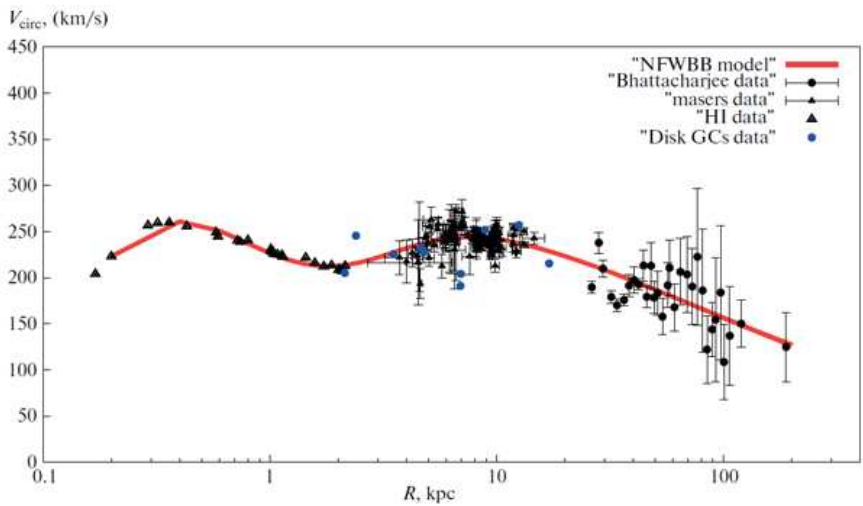}
\caption{Rotation curve corresponding to the NFWBB model of the
potential. The blue dots show the circular velocities of the disk
GCs with orbital eccentricities <0.2.} \label{rotc}
\end{center}
\end{figure*}

\subsection{Integration of the Orbits}
The equation of motion of a test particle in an axisymmetric
gravitational potential can be obtained from the Lagrangian of the
system $\pounds$ (see Appendix A in Irrgang et al. [14]):
\begin{equation}
 \begin{array}{lll}
 \pounds(R,Z,\dot{R},\dot{\psi},\dot{Z})=\\
 \qquad0.5(\dot{R}^2+(R\dot{\psi})^2+\dot{Z}^2)-\Phi(R,Z).
 \label{Lagr}
 \end{array}
\end{equation}
Introducing canonical moments
\begin{equation}
 \begin{array}{lll}
    p_{R}=\partial\pounds/\partial\dot{R}=\dot{R},\\
 p_{\psi}=\partial\pounds/\partial\dot{\phi}=R^2\dot{\psi},\\
    p_{Z}=\partial\pounds/\partial\dot{Z}=\dot{Z},
 \label{moments}
 \end{array}
\end{equation}
we obtain the Lagrange equations as a system of six first-order
differential equations: \begin{equation}
 \begin{array}{llllll}
 \dot{R}=p_R,\\
 \dot{\psi}=p_{\psi}/R^2,\\
 \dot{Z}=p_Z,\\
 \dot{p_R}=-\partial\Phi(R,Z)/\partial R +p_{\psi}^2/R^3,\\
 \dot{p_{\psi}}=0,\\
 \dot{p_Z}=-\partial\Phi(R,Z)/\partial Z.
 \label{eq-motion}
 \end{array}
\end{equation}
To integrate Eqs. (4), we used our own integrator that implements
the fourth-order Runge–Kutta algorithm.

The peculiar velocity of the Sun relative to the local standard of
rest was considered equal to
$(u_\odot,v_\odot,w_\odot)=(11.1,12.2,7.3)\pm(0.7,0.5,0.4)$ km
s$^{-1}$ [17]. Here, we use heliocentric velocities in a moving
Cartesian coordinate system with $u$ velocity directed toward the
galactic center, $v$ in the direction of the Galaxy rotation, and
$w$ perpendicular to the plane of the Galaxy and directed to the
north pole of the Galaxy.

et the initial positions and spatial velocities of the test
particle in the heliocentric coordinate system be
$(x_o,y_o,z_o,u_o,v_o,w_o)$. The initial positions ($X,Y,Z$) and
velocities ($U,V,W$) of the test particle in the Cartesian
coordinates of the Galaxy are given by the formulas
\begin{equation}
 \begin{array}{llllll}
 X=R_\odot-x_o, Y=y_o, Z=z_o+h_\odot,\\
 R=\sqrt{X^2+Y^2},\\
 U=u_o+u_\odot,\\
 V=v_o+v_\odot+V_\odot,\\
 W=w_o+w_\odot,
 \label{init}
 \end{array}
\end{equation}
where $R_\odot$ and $V_\odot$ are the galactocentric distance and
linear velocity of rotation of the local standard of rest around
the center of the Galaxy, and $h_\odot=16$~pc [18] is the
elevation of the Sun above the plane of the Galaxy.

In this study, we calculate the following orbital parameters of
globular clusters: (1) the initial distance of the GC from the
center of the Galaxy $d_{GC}$; (2) radial velocity $\Pi$; (3)
circular velocity $\Theta$; (4) total 3D velocity $V_{tot}$; (5)
apocentric distance $(apo)$ of the orbit; (6) pericentric distance
$(peri)$ of the orbit; (7) eccentricity $(ecc)$ of the orbit; (8)
angular momentum components; (9) the angle of orbital inclination
$\theta$; (10) orbital period $T_r$; and (11) total energy $E$.

The formulas for calculating all of the listed orbital parameters
are given in [7]. The uncertainties of the orbital parameters were
calculated by the Monte Carlo method using 100 iterations
considering the uncertainties in the initial GC coordinates and
velocities, as well as errors in the peculiar velocity of the Sun.

\section{DATA}
For the previously examined 152 globular clusters [7] with the
data mainly from Vasiliev’s catalog [3], we took only the new
values of mean proper motions and their uncertainties from the new
catalog by Vasiliev and Baumgardt [8] obtained from the Gaia EDR3
catalog data. All other astrometric data (distances, radial
velocities, coordinates) remained the same. Although the new
catalog of Vasiliev and Baumgardt [8] also contains new, more
accurate mean values of trigonometric parallaxes, their
uncertainty remains rather large as compared to the distances
found from the horizontal giant branch [19] in Vasiliev’s catalog
[3].

In Fig. 2, we give a comparison of the mean proper motions from
these two catalogs obtained from the Gaia DR2 and Gaia EDR3
measurements. As follows from the figure, the new proper motion
values for a number of GCs differ markedly from the old values. At
the same time, the accuracy of measuring the new proper motions
has doubled on average.

\begin{figure*}
\begin{center}
   \includegraphics[width=0.7\textwidth]{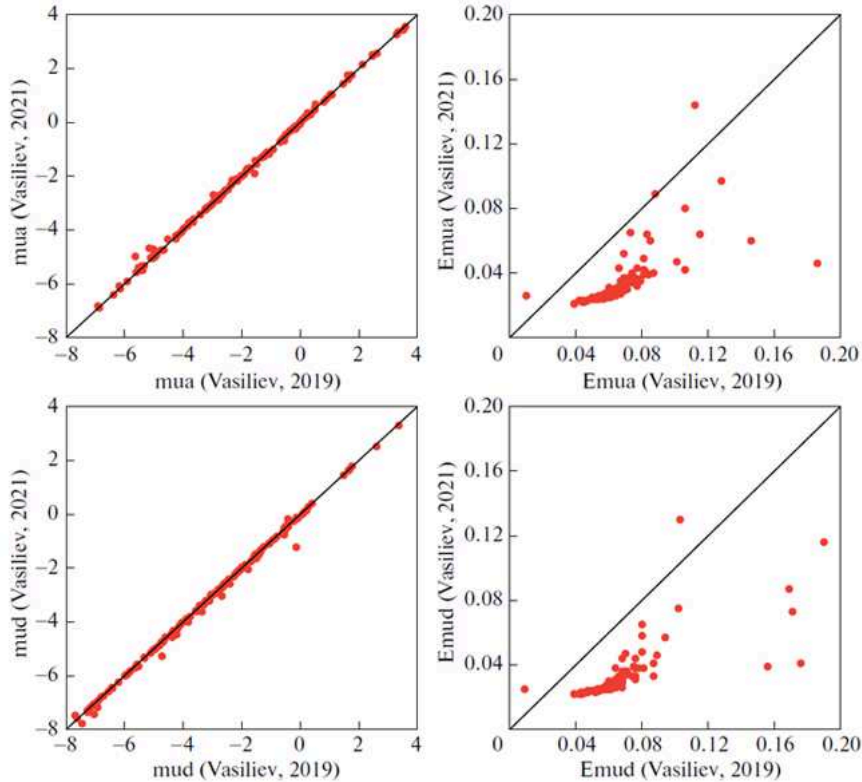}
\caption{Comparison of the GC proper motions (by $\alpha$, mua and
$\delta$, mud) and their uncertainties (Emua and Emud,
respectively) from Vasiliev’s catalog [3] (Gaia DR2, horizontal
axis) and Vasiliev and Baumgardt’s catalog [8] (Gaia EDR3,
vertical axis). Each panel has a coincidence line.}
\end{center}
\end{figure*}

\begin{figure*}
\begin{center}
   \includegraphics[width=0.95\textwidth]{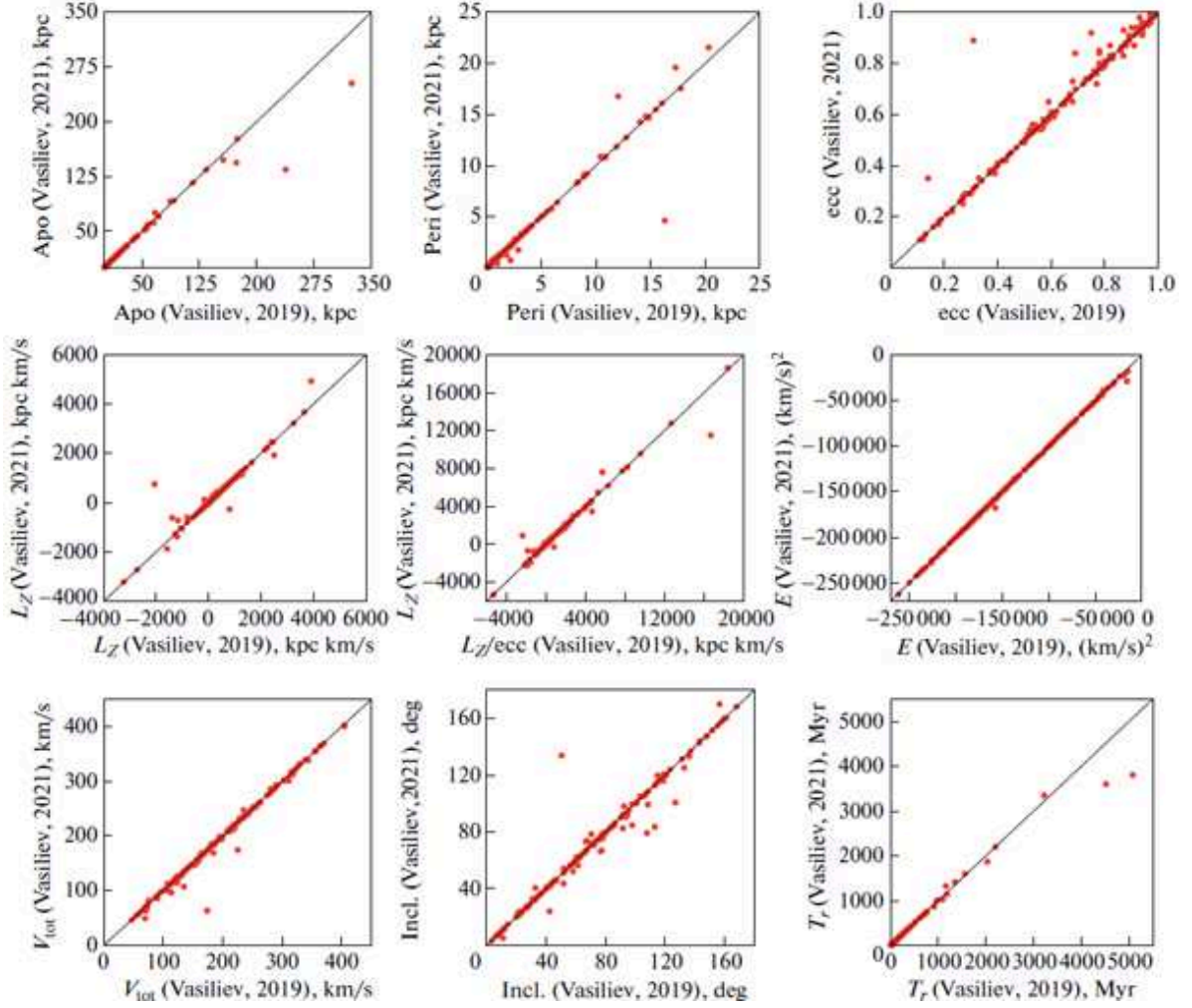}
\caption{Comparison of the GC orbital parameters ($apo, peri, ecc,
L_z, l_z/ecc, E, V_{tot}, \theta, T_r$) obtained using Vasiliev’s
catalog (2019) [3] with mean proper motions from Gaia DR2
(horizontal axis) and Vasiliev and Baumgardt’s catalog (2021) [8]
with mean proper motions from Gaia EDR3 (vertical axis). Each
panel has a diagonal coincidence line.}
\end{center}
\end{figure*}

\begin{figure*}
\begin{center}
   \includegraphics[width=0.95\textwidth]{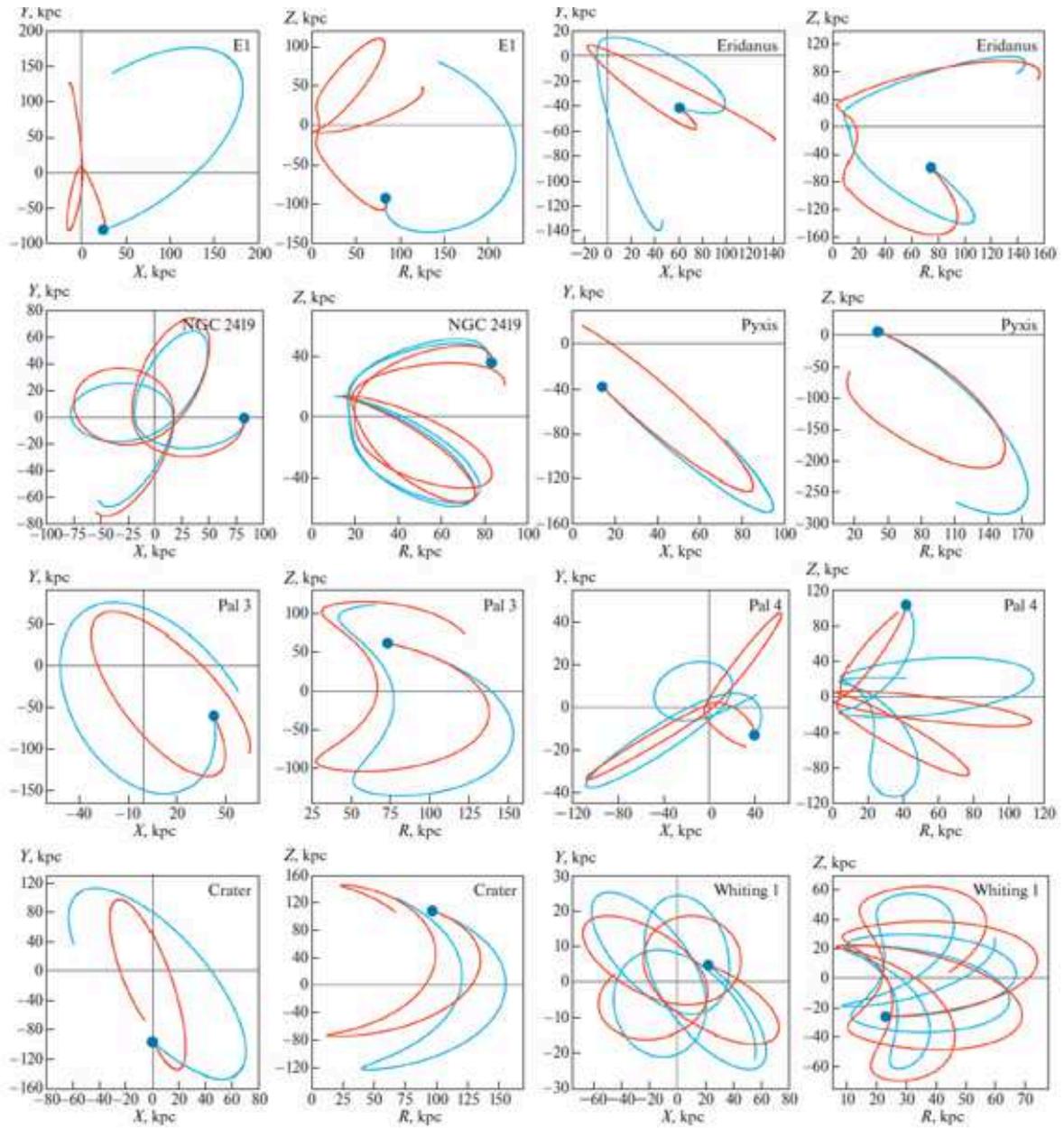}
\caption{Orbits of very distant GCs. Blue color shows GC orbits
with proper motions from Gaia DR2, red color from Gaia EDR3. The
beginning of the orbit is marked with a blue circle.}
\end{center}
\end{figure*}

\begin{figure*}
\begin{center}
   \includegraphics[width=0.85\textwidth]{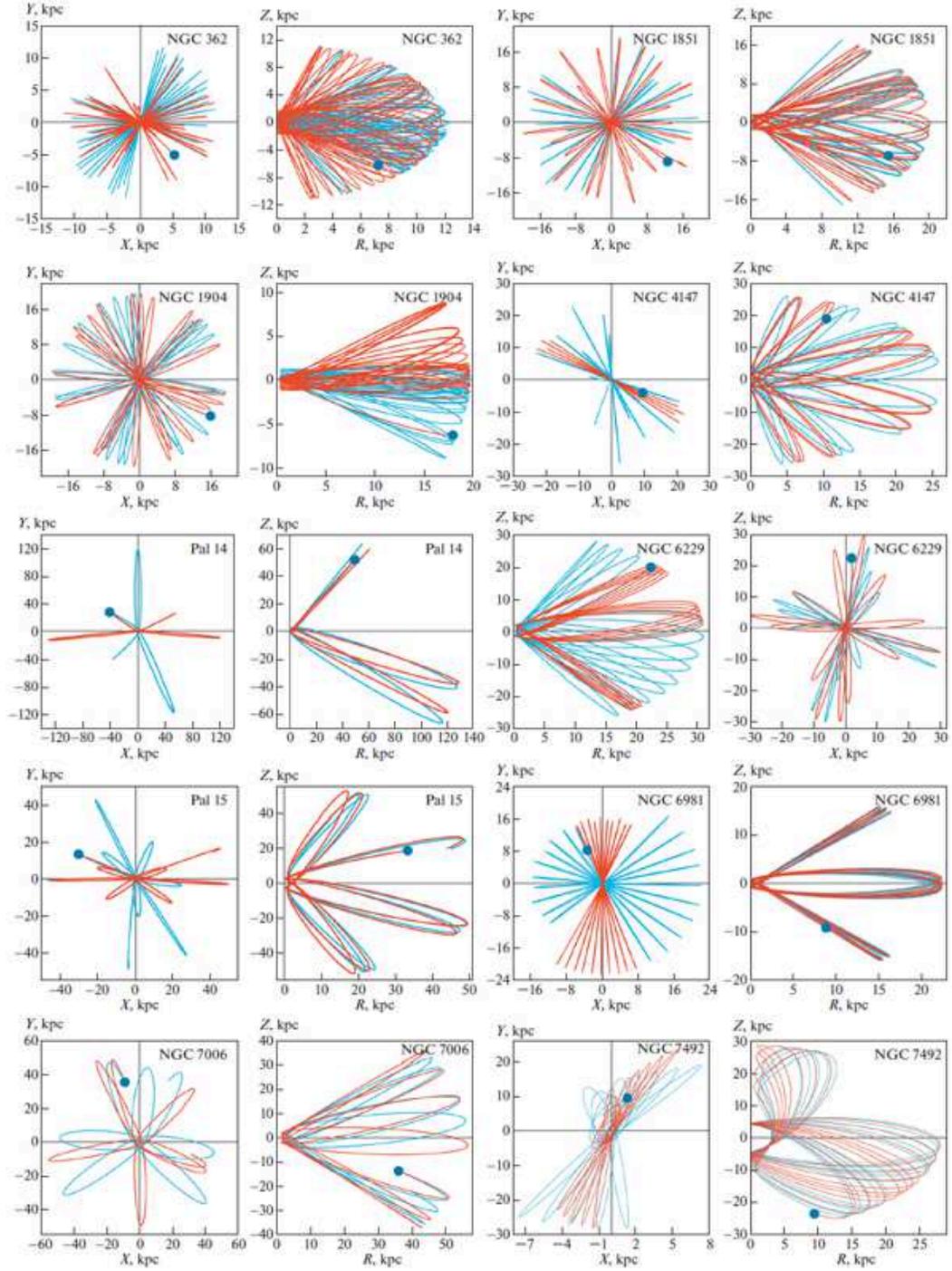}
\caption{Strongly radially elongated GC orbits. Blue color shows
GC orbits with proper motions from Gaia DR2, red color from Gaia
EDR3. The beginning of the orbit is marked with a blue circle.}
\end{center}
\end{figure*}

\begin{figure*}
\begin{center}
   \includegraphics[width=0.7\textwidth]{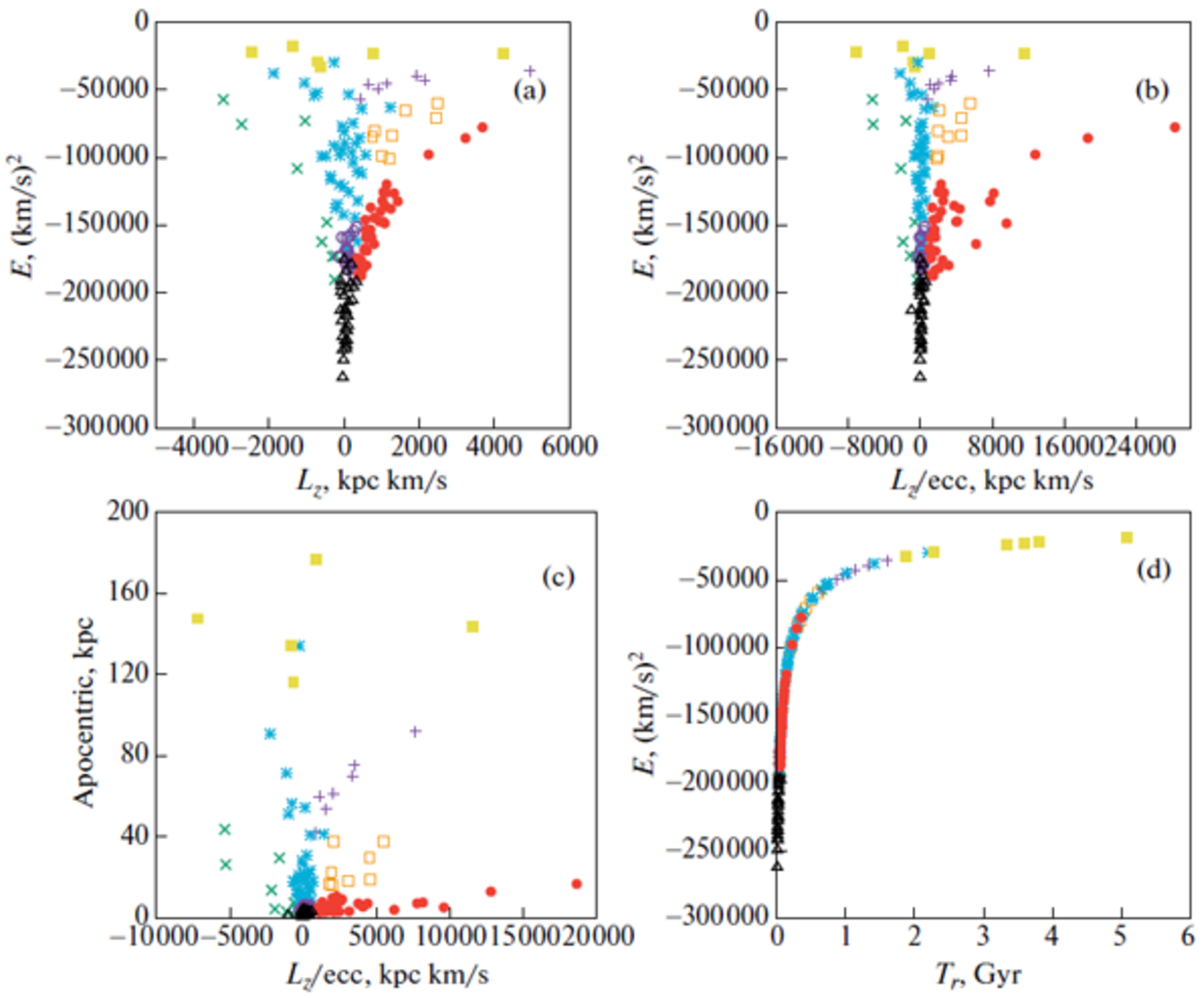}
\caption{Two-dimensional diagrams (a) $L_Z-E$, (b) $L_Z/ecc-E$,
(c) $L_Z/ecc-apo$, and (d) $T_r-E$. Different symbols indicate the
belonging of the GC to different subsystems of the Galaxy: red
circles—disk (D), black triangles—bulge (B), blue stars—Gaia-
Enceladus (GE), orange squares—Helmi stream (H99), green
crosses—Sequoia galaxy (Seq), purple crosses—Sagittarius dwarf
galaxy (Sgr), yellow squares—unassociated high energy group (HE),
purple open circles—unassociated low energy group (LE).}
\end{center}
\end{figure*}

\section{RESULTS}
Table 1 shows the orbital parameters of 152 GCs listed in Section
2.2, which are calculated for the new mean proper motions from
EDR3 [8]. 

The values of the parameters obtained from the integration of the
orbits 5 Gyr back are given for each GC. The columns show the
initial GC distance from the center of the Galaxy $d_{GC}$; radial
velocity $\Pi$; circular velocity $\Theta$; total 3D velocity $V_{tot}$;
apocentric distance (apo) of the orbit; pericentric distance
(peri) of the orbit; eccentricity (ecc) of the orbit; orbital
inclination angle $\theta$; orbital period $T_r$; $Z$ component of
the angular momentum $L_Z$; total energy $E$; and subsystem in the
Galaxy to which the object is assigned. The
objects with the greatest changes in the orbital properties are
marked with an asterisk.

The column  shows the
belonging of the clusters to a subsystem of the Galaxy. The
following designations of subsystems are adopted, similar to those
in [5]: D is disk; B is bulge; GE is the Sausage galaxy, or
Gaia–Enceladus; H99 is the Helmi stream; Seq is the Sequoia
galaxy; Sgr is the Sagittarius dwarf galaxy; HE is the
unassociated high energy group; and LE is the unassociated low
energy group. The GCs were divided into subsystems of the bulge
(B), thick disk (D), and halo in accordance with the algorithm we
proposed [4]. The GC classification by halo subsystems (Seq, Sqr,
GE, H99, HE, LE) was proposed by Massari et al. [5]. In [7], we
modified the classification of Massari et al. in accordance with
the orbital features of the GCs. Twenty-seven objects were
modified. The analysis of the new orbital parameters kept this
classification in force.

The GC orbital parameters calculated using the Gaia EDR3 mean
proper motions were compared with the similar parameters [7]
calculated from the Gaia DR2 data, which is shown in Fig. 3. As
can be seen from the figures, a significant difference is observed
for a number of GCs. We were able to identify 22 most prominent
representatives.

As the analysis showed, the greatest change in the orbital
parameters was experienced by very distant GCs, the measurement of
proper motions associated with the greatest difficulties. Due to
the increased accuracy of the data from the Gaia EDR3 catalog in
comparison with the data from Gaia DR2, the proper motion values
for distant objects have changed greatly. These objects are E1,
Eridanus, Pyxis, Palomar 3, Palomar 4, and Crater belonging to the
HE subsystem, and NGC 2419 and Whiting 1 belonging to Sqr. The
orbits of these GCs in two projections $(X,Y)$ and $(R,Z)$ were
plotted from the EDR3 and DR2 data are shown in Fig. 4 in red and
blue, respectively. As can be seen from the figures, the greatest
change in the orbit affected the most distant object E1.

Another class of objects that have undergone significant changes
in orbital motion are GCs with highly elongated radial orbits. We
have separately identified 10 most prominent representatives: NGC
362, NGC 1851, NGC 1904, NGC 4147, Palomar 14, NGC 6229, Palomar
15, NGC 6981, NGC 7006, and NGC 7492. They all belong to the GE
subsystem. The orbits of these globular clusters are shown in Fig.
5. As can be seen from the figure, the orbits of three GCs
underwent the most significant changes: Palomar 14, which is also
among the most distant objects, Palomar 15, and NGC 6981.

It should be noted that the clusters E1, Eridanus, Pyxis, Palomar
3, Palomar 4, Crater, and Palomar 14 gave the largest deviations
from the coincidence line in Fig. 3.

Four other GCs whose orbits have changed not so much but still
noticeably are NGC 5946 and FSR 1735 of the LE subsystem, ESO
280--06 (GE), and the disk object NGC 6569.

Thus, the most distant objects and objects with orbits strongly
elongated in the radial direction have undergone the greatest
change in the orbital properties. All 22 GCs with significantly
changed orbital properties are marked by an asterisk in the last
column of Table 1. The orbits of the other GCs have changed very
insignificantly, so the catalog of orbits presented in [7] can be
used to study their dynamics.

Figure 6 shows two-dimensional diagrams  (a) $L_Z-E$, (b)
$L_Z/ecc-E$, (c) $L_Z/ecc-apo$, and (d) $T_r-E$. Diagrams (a),
(b), and (c) are interesting in that they clearly show the
fragmentation of the GC distribution over the subsystems of the
Galaxy. Diagram (b) with the $L_Z/ecc$ values plotted along the
horizontal axis instead of $L_Z$ in comparison with diagram (a)
allows for a finer structuring of the distribution of the thick
disk GCs (D). This is due to the fact that disk objects have
relatively small values of eccentricity, which allows
``spreading'' them along the abscissa axis as a result of division
by $ecc$. The fine structure of the disk object distribution is
also clearly visible in diagram (c). Apparently, further study of
the structural distribution of GCs is of interest from the point
of view of a more subtle classification of GCs.

The $T_r-E$ diagram (d) is interesting in that it represents an
unambiguous relationship between the period and the total energy
of objects in the Galaxy set only by the gravitational potential
of the Galaxy. This dependence can be fairly accurately fitted by
a hyperbolic function. Thus, for example, knowing the total energy
of a GC, one can immediately estimate the period of the orbit
without integrating it. Conversely, knowing the period of the
orbit, one can estimate the total energy $E$ of a GC.

\section{CONCLUSIONS}
The emergence of increasingly more accurate astrometric data on
the coordinates and spatial velocities of globular clusters makes
it possible to study their motion in three-dimensional space by
integrating the orbits in the gravitational potential of the
Galaxy.

Thanks to the Gaia DR2 data [1–3] on the proper motions of almost
all globular clusters known to date, it became possible to study
their kinematics and dynamics and classify GCs according to the
subsystems of the Milky Way in order to identify objects that
formed directly in the Galaxy or were brought from outside as a
result of accretion from other (dwarf) galaxies around the Milky
Way. The creation of the catalog of orbits and their parameters
for more than 150 GCs [7] with known data on the 6d phase space
required for the integration of the orbits provides highly
informative material for further research.

The recent appearance of a new, more accurate version of the
catalog of proper motions posed the task of specifying the orbital
motion of the GCs, which was the subject of this study. As a
result of using new, more accurate proper motions determined from
the Gaia EDR3 data, we were able to identify 22 GCs whose orbital
motion underwent a significant change. Those are NGC 362, Whiting
1, E1, Eridanus, NGC 1851, NGC 1904, NGC 2419, Pyxis, Palomar 3,
Palomar 4, Crater, NGC 4147, NGC 5946, Palomar 14, NGC 6229, FSR
1735, Palomar 15, ESO 280– 06, NGC 6569, NGC 6981, NGC 7006, and
NGC 7492. The major part of this list are very distant GCs, as
well as GCs with orbits strongly elongated toward the center of
the Galaxy, belonging to the Gaia-Enceladus subsystem.

Thus, the main result of this study is the creation of a new
catalog of orbital parameters of 152 GCs with proper motions from
Gaia EDR3, indicating objects with significantly changed orbital
properties in comparison with the catalog based on the proper
motions from Gaia DR2.

\subsubsection*{ACKNOWLEDGMENTS}
The authors are grateful to the anonymous reviewer for comments
that have helped improve the article.

 \medskip\subsubsection*{REFERENCES}

 {\small
 \quad
 ~1. A. Helmi, F. van Leeuwen, P. J. McMillan, D. Massari,
et al., Astron. Astrophys. 616, A12 (2018).

2. H. Baumgardt, M. Hilker, A. Sollima, and A. Bellini, Mon. Not.
R. Astron. Soc. 482, 5138 (2019).

3. E. Vasiliev, Mon. Not. R. Astron. Soc. 484, 2832 (2019).

4. A. T. Bajkova, G. Carraro, V. I. Korchagin, N. O. Budanova, and
V. V. Bobylev, Astrophys. J. 895, 69 (2020).

5. D. Massari, H. H. Koppelman and A. Helmi, Astron. Astrophys.
630, L4 (2019).

6. G. C. Myeong, E. Vasiliev, G. Iorio, N. W. Evans, and V.
Belokurov, Mon. Not. R. Astron. Soc. 488, 1235 (2019).

 7. A. T. Bajkova and V.V. Bobylev, Res. Astron. Astrophys. {\bf 21}, 173 (2021).

8. E. Vasiliev and H. Baumgardt, arXiv: 2102.09568 [astro-ph.GA]
(2021).

9. A. T. Bajkova and V. V. Bobylev, Astron. Lett. 42, 567 (2016).

10. A. T. Bajkova and V. V. Bobylev, Open Astron. 26, 72 (2017).

11. M. Miyamoto and R. Nagai, Publ. Astron. Soc. Jpn. 27, 533
(1975).

12. J. F. Navarro, C. S. Frenk, and S. D. M. White, Astrophys. J.
490, 493 (1997).

13. P. Bhattacharjee, S. Chaudhury, and S. Kundu, Astrophys. J.
785, 63 (2014).

14. A. Irrgang, B. Wilcox, E. Tucker, and L. Schiefelbein, Astron.
Astrophys. 549, A137 (2013).

15. H. H. Koppelman and A. Helmi, arXiv: 2006.16283 [astro-ph.GA]
(2020).

16. W. Wang, J. Han, M. Cautun, Z. Li, and M. Ishigaki, Sci. China
Phys. Mech. Astron. 63, 109801 (2020).

17. R. Sch\"onrich, J. Binney, and W. Dehnen, Mon. Not. R. Astron.
Soc. 403, 1829 (2010).

18. V. V. Bobylev and A. T. Bajkova, Astron. Lett. 42, 1 (2016).

19. W. Harris, arXiv: 1012.3224 [astro-ph.GA] (2010).
 }

\rotatebox{90}{
        \begin{minipage}{1.5\linewidth}
        \begin{scriptsize}
        \begin{center}
    Table 1. Orbital properties of GCs    \medskip
        \begin{tabular}{|l|r|r|r|r|r|r|r|r|r|r|r|r|r|}\hline
 Name &$d_{GC}$&$\Pi$ &$\Theta$ &$V_{tot}$ &apo   & peri      &ecc&incl.    &$T_r$&$L_Z$&$E$ &Gal.& \\
      & [kpc]  &[km/s]&[km/s]   &[km/s]    & [kpc]& [kpc]&   & $\theta$&[Gyr]&[kpc  &[km$^2$ &Sub-&\\
      &        &      &         &          &      &      &   & [deg]   &     & km/s]&/s$^2$&syst&\\\hline
NGC 104    &   7.6  & $    6^{+  6}_{-  5}$& $  191^{+  4}_{-  4}$& $  197^{+  4}_{-  4}$& $  7.7^{+0.1}_{-0.1}$& $ 5.52^{+0.22}_{-0.19}$& $0.16^{+0.02}_{-0.01}$& $  28^{+   1}_{-   1}$& $ 116^{+   3}_{-   3}$& $   1325^{+  34}_{-  36}$& $   -126360^{+   1237}_{-   1071}$& D & \\
NGC 288    &  12.2  & $    4^{+  1}_{-  2}$& $  -42^{+ 17}_{- 13}$& $   66^{+  9}_{- 10}$& $ 12.4^{+0.3}_{-0.4}$& $ 1.33^{+0.45}_{-0.53}$& $0.81^{+0.07}_{-0.06}$& $ 121^{+   4}_{-   8}$& $ 142^{+   5}_{-   5}$& $   -349^{+ 142}_{- 105}$& $   -116316^{+   1654}_{-   1780}$&GE &\\
NGC362    &   9.5  & $  126^{+  5}_{-  9}$& $   -2^{+  7}_{- 11}$& $  144^{+  5}_{-  8}$& $ 11.7^{+0.5}_{-0.2}$& $ 0.08^{+0.22}_{-0.00}$& $0.99^{+0.00}_{-0.04}$& $  93^{+  18}_{-  10}$& $ 126^{+   5}_{-   3}$& $    -12^{+  48}_{-  79}$& $   -121469^{+   1964}_{-   1664}$&GE &*\\
Whiting 1    &  34.7  & $ -232^{+ 11}_{- 16}$& $   85^{+ 13}_{- 15}$& $  247^{+ 16}_{- 10}$& $ 75.4^{+17.0}_{-9.0}$& $21.56^{+1.86}_{-1.78}$& $0.56^{+0.05}_{-0.03}$& $  74^{+   3}_{-   2}$& $1328^{+ 339}_{- 177}$& $   1927^{+ 305}_{- 362}$& $    -39308^{+   4454}_{-   3011}$&Sgr &*\\
NGC 1261   &  18.2  & $  -98^{+  5}_{-  8}$& $  -21^{+  7}_{-  6}$& $  121^{+  7}_{-  5}$& $ 21.2^{+0.8}_{-0.6}$& $ 0.81^{+0.35}_{-0.14}$& $0.93^{+0.01}_{-0.03}$& $ 120^{+   7}_{-  11}$& $ 246^{+  10}_{-   9}$& $   -273^{+  95}_{-  82}$& $    -91128^{+   1908}_{-   1335}$&GE &\\
Pal 1     &  17.4  & $   43^{+  5}_{-  6}$& $  215^{+  2}_{-  2}$& $  221^{+  2}_{-  2}$& $ 19.2^{+0.5}_{-0.6}$& $14.83^{+0.47}_{-0.44}$& $0.13^{+0.02}_{-0.02}$& $  15^{+   1}_{-   0}$& $ 356^{+  10}_{-  10}$& $   3677^{+  69}_{-  82}$& $    -77542^{+    946}_{-   1095}$&D &\\
E 1       & 124.7  & $  -16^{+ 58}_{- 36}$& $   -8^{+ 52}_{- 56}$& $   63^{+ 78}_{-  0}$& $134.5^{+64.9}_{-24.9}$& $ 8.03^{+63.43}_{-0.00}$& $0.89^{+0.00}_{-0.49}$& $ 101^{+  22}_{-  32}$& $2280^{+1993}_{- 150}$& $   -706^{+4263}_{-4646}$& $    -28995^{+   8306}_{-    453}$&HE &*\\
Eridanus    &  95.2  & $  -74^{+ 18}_{- 14}$& $   10^{+ 18}_{- 12}$& $  166^{+ 12}_{- 19}$& $176.7^{+20.3}_{-27.8}$& $16.77^{+5.66}_{-4.38}$& $0.83^{+0.03}_{-0.06}$& $  84^{+   7}_{-  13}$& $3348^{+ 524}_{- 641}$& $    753^{+1356}_{- 874}$& $    -23513^{+   1835}_{-   2954}$&HE &*\\
Pal 2     &  35.3  & $ -107^{+  3}_{-  4}$& $   13^{+  9}_{- 13}$& $  108^{+  4}_{-  2}$& $ 40.9^{+1.7}_{-1.8}$& $ 0.84^{+0.54}_{-0.46}$& $0.96^{+0.02}_{-0.03}$& $  24^{+  66}_{-  21}$& $ 506^{+  26}_{-  24}$& $    467^{+ 320}_{- 446}$& $    -63898^{+   1598}_{-   1704}$&GE &\\
NGC 1851   &  16.9  & $  104^{+  2}_{-  3}$& $   -5^{+  4}_{-  4}$& $  132^{+  4}_{-  3}$& $ 20.1^{+0.4}_{-0.4}$& $ 0.14^{+0.23}_{-0.00}$& $0.99^{+0.00}_{-0.03}$& $  98^{+   7}_{-   7}$& $ 228^{+   7}_{-   4}$& $    -76^{+  66}_{-  60}$& $    -94253^{+   1069}_{-   1078}$& GE&*\\
NGC 1904   &  19.0  & $   45^{+  4}_{-  4}$& $   11^{+  5}_{-  6}$& $   47^{+  4}_{-  3}$& $ 19.7^{+0.2}_{-0.6}$& $ 0.35^{+0.25}_{-0.12}$& $0.97^{+0.01}_{-0.03}$& $  62^{+  16}_{-  14}$& $ 218^{+   8}_{-  10}$& $    192^{+  98}_{- 115}$& $    -96229^{+   1108}_{-    974}$&GE&*\\
NGC 2298   &  16.0  & $  -92^{+  7}_{-  6}$& $  -31^{+  6}_{-  6}$& $  124^{+  7}_{-  6}$& $ 18.0^{+0.4}_{-0.4}$& $ 1.25^{+0.34}_{-0.15}$& $0.87^{+0.01}_{-0.03}$& $ 117^{+   5}_{-   5}$& $ 208^{+   6}_{-   6}$& $   -494^{+  94}_{-  95}$& $    -98729^{+   1300}_{-   1171}$&GE & \\
NGC 2419   &  90.2  & $   -7^{+  6}_{-  5}$& $   59^{+ 13}_{- 13}$& $   80^{+ 14}_{- 10}$& $ 91.9^{+3.5}_{-3.2}$& $19.60^{+9.62}_{-5.54}$& $0.65^{+0.07}_{-0.12}$& $  44^{+   8}_{-   6}$& $1600^{+ 219}_{- 815}$& $   4937^{+1125}_{-1132}$& $    -35478^{+   1537}_{-   1190}$&Sgr &*\\
Pyxis     &  41.5  & $ -246^{+  4}_{-  4}$& $  -34^{+ 11}_{- 10}$& $  301^{+  4}_{-  6}$& $252.7^{+59.9}_{-43.8}$& $41.69^{+1.53}_{-20.48}$& $0.72^{+0.12}_{-0.00}$& $ 100^{+   3}_{-   3}$& $5084^{+1247}_{-1393}$& $  -1382^{+ 479}_{- 458}$& $    -18116^{+   2139}_{-   1970}$&HE &*\\
NGC 2808   &  11.3  & $ -157^{+  2}_{-  1}$& $   41^{+  4}_{-  4}$& $  165^{+  1}_{-  1}$& $ 14.4^{+0.5}_{-0.5}$& $ 1.01^{+0.09}_{-0.07}$& $0.87^{+0.01}_{-0.02}$& $  10^{+   2}_{-   1}$& $ 158^{+   6}_{-   6}$& $    460^{+  34}_{-  30}$& $   -111826^{+   1885}_{-   1764}$&GE &\\
E 3       &   9.3  & $   44^{+ 12}_{- 11}$& $  252^{+  8}_{-  9}$& $  275^{+  9}_{-  9}$& $ 13.1^{+1.3}_{-1.1}$& $ 9.18^{+0.30}_{-0.31}$& $0.18^{+0.04}_{-0.04}$& $  29^{+   1}_{-   1}$& $ 224^{+  14}_{-  12}$& $   2244^{+ 107}_{- 112}$& $    -97762^{+   3113}_{-   2858}$&D &\\
Pal 3     &  95.9  & $ -146^{+ 15}_{- 20}$& $   58^{+ 27}_{- 30}$& $  169^{+ 22}_{- 16}$& $144.0^{+59.4}_{-22.6}$& $66.53^{+14.11}_{-15.05}$& $0.37^{+0.12}_{-0.03}$& $  73^{+   8}_{-   7}$& $3602^{+1259}_{- 681}$& $   4236^{+2029}_{-2215}$& $    -22836^{+   4004}_{-   2698}$&HE &*\\
NGC 3201   &   9.1  & $ -114^{+  8}_{- 10}$& $ -301^{+  6}_{-  5}$& $  356^{+  7}_{-  6}$& $ 26.4^{+2.0}_{-1.8}$& $ 8.39^{+0.17}_{-0.23}$& $0.52^{+0.02}_{-0.03}$& $ 152^{+   1}_{-   1}$& $ 376^{+  25}_{-  26}$& $  -2727^{+  83}_{-  70}$& $    -75366^{+   2504}_{-   2517}$&Seq &\\
Pal 4     & 111.4  & $    2^{+ 19}_{- 21}$& $  -15^{+ 27}_{- 19}$& $   49^{+ 17}_{-  3}$& $116.4^{+0.0}_{-9.4}$& $ 4.61^{+9.33}_{-2.03}$& $0.92^{+0.03}_{-0.14}$& $ 105^{+   4}_{-  24}$& $1870^{+   0}_{- 620}$& $   -615^{+1107}_{- 807}$& $    -32274^{+   1573}_{-    650}$&HE &*\\
Crater    & 144.8  & $  -89^{+ 76}_{- 54}$& $  -25^{+ 64}_{-117}$& $  107^{+101}_{- 15}$& $147.7^{+426.9}_{-71.0}$& $71.90^{+84.04}_{-9.28}$& $0.35^{+0.33}_{-0.11}$& $  99^{+  28}_{-  19}$& $3806^{+1823}_{-3363}$& $  -2469^{+6352}_{-11393}$& $    -22149^{+  17069}_{-   2215}$&HE &*\\
NGC 4147   &  21.5  & $   47^{+  6}_{-  5}$& $   -0^{+ 12}_{- 13}$& $  136^{+  4}_{-  3}$& $ 26.4^{+1.1}_{-1.1}$& $ 0.39^{+0.55}_{-0.00}$& $0.97^{+0.00}_{-0.04}$& $  91^{+  15}_{-  12}$& $ 314^{+  15}_{-  14}$& $     -4^{+ 125}_{- 130}$& $    -81056^{+   1755}_{-   1876}$&GE &*\\
NGC 4372   &   7.3  & $   17^{+  7}_{-  8}$& $  133^{+  7}_{-  7}$& $  149^{+  7}_{-  7}$& $  7.3^{+0.3}_{-0.1}$& $ 3.00^{+0.24}_{-0.24}$& $0.42^{+0.03}_{-0.03}$& $  28^{+   1}_{-   2}$& $  98^{+   3}_{-   3}$& $    965^{+  61}_{-  54}$& $   -139619^{+   2139}_{-   1768}$&D &\\
Rup 106    &  18.5  & $ -243^{+  3}_{-  5}$& $   91^{+  8}_{- 10}$& $  261^{+  5}_{-  3}$& $ 38.0^{+2.2}_{-2.1}$& $ 4.65^{+0.51}_{-0.36}$& $0.78^{+0.02}_{-0.02}$& $  46^{+   4}_{-   3}$& $ 498^{+  34}_{-  32}$& $   1627^{+ 167}_{- 203}$& $    -64947^{+   2121}_{-   2229}$&H99 &\\
NGC 4590   &  10.3  & $ -169^{+  9}_{- 10}$& $  294^{+  8}_{- 10}$& $  339^{+  4}_{-  5}$& $ 30.0^{+1.7}_{-1.5}$& $ 8.89^{+0.29}_{-0.28}$& $0.54^{+0.02}_{-0.01}$& $  41^{+   2}_{-   1}$& $ 430^{+  28}_{-  22}$& $   2454^{+  95}_{-  99}$& $    -70313^{+   2039}_{-   1905}$&H99 &\\
NGC 4833   &   7.2  & $  105^{+  8}_{- 10}$& $   40^{+  9}_{- 11}$& $  120^{+  9}_{- 11}$& $  8.1^{+0.3}_{-0.3}$& $ 0.74^{+0.14}_{-0.20}$& $0.83^{+0.04}_{-0.02}$& $  37^{+  10}_{-   6}$& $  86^{+   3}_{-   4}$& $    288^{+  65}_{-  83}$& $   -144343^{+   2052}_{-   2144}$&GE &\\
NGC 5024   &  18.5  & $  -95^{+  5}_{-  4}$& $  143^{+  4}_{-  6}$& $  186^{+  4}_{-  6}$& $ 22.4^{+0.9}_{-1.0}$& $ 9.11^{+0.51}_{-0.61}$& $0.42^{+0.02}_{-0.01}$& $  74^{+   1}_{-   1}$& $ 336^{+  16}_{-  19}$& $    810^{+  29}_{-  39}$& $    -79888^{+   1805}_{-   2220}$&H99 &\\
NGC 5053   &  17.9  & $  -91^{+  4}_{-  3}$& $  137^{+  6}_{-  6}$& $  168^{+  5}_{-  6}$& $ 18.0^{+0.9}_{-0.7}$& $10.87^{+0.97}_{-0.63}$& $0.25^{+0.02}_{-0.04}$& $  76^{+   1}_{-   1}$& $ 304^{+  18}_{-  15}$& $    747^{+  43}_{-  39}$& $    -84546^{+   2416}_{-   2020}$&H99 &\\
NGC 5139   &   6.6  & $  -70^{+  4}_{-  4}$& $  -72^{+  4}_{-  5}$& $  128^{+  6}_{-  4}$& $  7.4^{+0.0}_{-0.3}$& $ 1.13^{+0.16}_{-0.07}$& $0.74^{+0.01}_{-0.04}$& $ 137^{+   3}_{-   4}$& $  80^{+   4}_{-   1}$& $   -463^{+  31}_{-  34}$& $   -147822^{+   1693}_{-   1521}$&Seq &\\
NGC 5272   &  12.2  & $  -39^{+  4}_{-  4}$& $  142^{+  6}_{-  5}$& $  199^{+  4}_{-  4}$& $ 15.9^{+0.7}_{-0.6}$& $ 5.14^{+0.48}_{-0.21}$& $0.51^{+0.01}_{-0.03}$& $  57^{+   1}_{-   1}$& $ 210^{+  13}_{-   7}$& $    987^{+  44}_{-  39}$& $    -98373^{+   2147}_{-   2008}$&H99 &\\
NGC 5286   &   8.9  & $ -219^{+  2}_{-  2}$& $  -43^{+  8}_{-  8}$& $  223^{+  1}_{-  1}$& $ 13.7^{+0.6}_{-0.7}$& $ 0.80^{+0.20}_{-0.12}$& $0.89^{+0.01}_{-0.03}$& $ 124^{+   4}_{-   5}$& $ 152^{+   8}_{-   7}$& $   -372^{+  69}_{-  68}$& $   -113576^{+   2188}_{-   2745}$&H99 &\\
NGC 5466   &  16.4  & $  169^{+ 13}_{- 17}$& $ -137^{+ 15}_{- 16}$& $  313^{+  8}_{- 10}$& $ 51.3^{+5.3}_{-5.3}$& $ 5.76^{+0.66}_{-0.62}$& $0.80^{+0.01}_{-0.02}$& $ 108^{+   2}_{-   1}$& $ 712^{+  87}_{-  89}$& $   -795^{+  91}_{-  96}$& $    -54156^{+   3099}_{-   3610}$&GE &\\
NGC 5634   &  21.1  & $  -47^{+  6}_{-  8}$& $   34^{+ 10}_{- 12}$& $   63^{+  5}_{-  2}$& $ 21.6^{+0.9}_{-1.0}$& $ 2.05^{+0.36}_{-0.23}$& $0.83^{+0.01}_{-0.03}$& $  72^{+   6}_{-   4}$& $ 256^{+  14}_{-  11}$& $    303^{+  88}_{- 104}$& $    -89241^{+   2025}_{-   2042}$&GE &\\
NGC 5694   &  29.3  & $ -182^{+  6}_{-  5}$& $  -45^{+ 11}_{- 11}$& $  254^{+  7}_{-  7}$& $ 71.5^{+5.4}_{-4.8}$& $ 2.81^{+11.27}_{-2.36}$& $0.92^{+0.08}_{-0.20}$& $ 133^{+   4}_{-   7}$& $1012^{+  28}_{-  91}$& $  -1058^{+ 279}_{- 296}$& $    -44810^{+   2109}_{-   2055}$&GE &\\
IC 4499   &  15.6  & $ -243^{+  3}_{-  2}$& $  -73^{+  8}_{-  8}$& $  261^{+  2}_{-  2}$& $ 29.7^{+1.6}_{-1.8}$& $ 6.38^{+0.52}_{-0.20}$& $0.65^{+0.01}_{-0.03}$& $ 113^{+   2}_{-   2}$& $ 402^{+  23}_{-  25}$& $  -1042^{+ 118}_{- 116}$& $    -72691^{+   1994}_{-   2350}$&Seq &\\
NGC 5824   &  25.7  & $  -41^{+  9}_{- 11}$& $  109^{+ 12}_{- 12}$& $  216^{+ 10}_{-  8}$& $ 37.6^{+3.2}_{-2.2}$& $14.26^{+1.76}_{-1.73}$& $0.45^{+0.05}_{-0.04}$& $  58^{+   3}_{-   2}$& $ 602^{+  63}_{-  46}$& $   2464^{+ 297}_{- 298}$& $    -59518^{+   2892}_{-   2423}$&H99 &\\
Pal 5     &  18.4  & $  -53^{+  1}_{-  3}$& $  161^{+ 22}_{- 19}$& $  170^{+ 22}_{- 17}$& $ 18.9^{+1.1}_{-0.6}$& $10.77^{+2.65}_{-1.74}$& $0.27^{+0.08}_{-0.09}$& $  66^{+   3}_{-   3}$& $ 310^{+  42}_{-  26}$& $   1262^{+ 230}_{- 175}$& $    -83186^{+   4776}_{-   2994}$&H99 &\\
NGC 5897   &   7.3  & $   86^{+ 18}_{- 15}$& $   96^{+ 15}_{- 16}$& $  158^{+ 14}_{- 12}$& $  8.7^{+0.5}_{-0.3}$& $ 1.89^{+0.43}_{-0.35}$& $0.64^{+0.06}_{-0.05}$& $  61^{+   4}_{-   3}$& $ 106^{+   7}_{-   5}$& $    359^{+  83}_{-  78}$& $   -131869^{+   3454}_{-   2289}$& GE &\\
NGC 5904   &   6.3  & $ -291^{+ 14}_{- 10}$& $  126^{+ 10}_{-  8}$& $  365^{+ 11}_{- 13}$& $ 23.4^{+2.5}_{-2.6}$& $ 2.27^{+0.40}_{-0.08}$& $0.82^{+0.01}_{-0.03}$& $  72^{+   2}_{-   2}$& $ 288^{+  33}_{-  34}$& $    401^{+  32}_{-  41}$& $    -85416^{+   4305}_{-   4892}$& GE &\\
\hline
   \end{tabular}
   \end{center}
   \end{scriptsize}
   \end{minipage}
  }

 \rotatebox{90}{
        \begin{minipage}{1.5\linewidth}
        \begin{scriptsize}
        \begin{center}
        \begin{tabular}{|l|r|r|r|r|r|r|r|r|r|r|r|r|r|}\hline
 Name &$d_{GC}$&$\Pi$ &$\Theta$ &$V_{tot}$ &apo   & peri      &ecc&incl.    &$T_r$&$L_Z$&$E$ &Galactic& \\
      & [kpc]  &[km/s]&[km/s]   &[km/s]    & [kpc]& [kpc]&   & $\theta$&[Gyr]&[kpc  &[km$^2$ & Sub-&\\
      &        &      &         &          &      &      &   & [deg]   &     & km/s]&/s$^2$&system&\\\hline
NGC 5927   &   4.7  & $  -39^{+ 15}_{- 15}$& $  233^{+  5}_{-  8}$& $  236^{+  6}_{-  8}$& $  5.2^{+0.3}_{-0.3}$& $ 4.17^{+0.28}_{-0.29}$& $0.11^{+0.05}_{-0.03}$& $   9^{+   1}_{-   1}$& $  82^{+   5}_{-   4}$& $   1077^{+  62}_{-  61}$& $   -148646^{+   3031}_{-   2848}$&D &\\
NGC 5946   &   5.8  & $   36^{+ 13}_{-  9}$& $   25^{+  8}_{-  7}$& $  115^{+  8}_{-  4}$& $  5.9^{+0.3}_{-0.2}$& $ 0.35^{+0.19}_{-0.05}$& $0.89^{+0.01}_{-0.05}$& $  76^{+   4}_{-   4}$& $  66^{+   3}_{-   4}$& $    145^{+  49}_{-  42}$& $   -158142^{+   2872}_{-   2154}$&LE &*\\
ESO 224-8   &  12.6  & $  -44^{+ 22}_{- 22}$& $  257^{+ 17}_{- 19}$& $  261^{+ 17}_{- 17}$& $ 16.8^{+3.0}_{-2.4}$& $11.84^{+0.60}_{-1.30}$& $0.17^{+0.08}_{-0.04}$& $   7^{+   1}_{-   0}$& $ 290^{+  40}_{-  40}$& $   3226^{+ 238}_{- 333}$& $    -85573^{+   5016}_{-   6103}$&D &\\
NGC 5986   &   4.7  & $   61^{+ 15}_{- 12}$& $   23^{+  7}_{-  9}$& $   66^{+ 14}_{- 10}$& $  5.5^{+0.1}_{-0.6}$& $ 0.16^{+0.11}_{-0.05}$& $0.94^{+0.02}_{-0.04}$& $  65^{+  10}_{-   6}$& $  56^{+   2}_{-   4}$& $     94^{+  30}_{-  39}$& $   -168587^{+   2093}_{-   2777}$&LE &\\
FSR 1716   &   4.8  & $   87^{+ 22}_{- 28}$& $  228^{+  9}_{- 14}$& $  286^{+  7}_{- 12}$& $  7.2^{+0.5}_{-0.7}$& $ 3.96^{+0.37}_{-0.38}$& $0.29^{+0.06}_{-0.07}$& $  33^{+   2}_{-   2}$& $ 108^{+   6}_{-   6}$& $   1089^{+  62}_{-  83}$& $   -135629^{+   3132}_{-   4361}$&D &\\
Pal 14    &  71.4  & $  123^{+ 11}_{- 11}$& $   -6^{+ 18}_{-  6}$& $  177^{+ 11}_{- 12}$& $134.2^{+9.8}_{-12.9}$& $ 0.73^{+2.00}_{-0.02}$& $0.99^{+0.00}_{-0.03}$& $ 134^{+   0}_{-  84}$& $2200^{+ 218}_{- 263}$& $   -273^{+ 907}_{- 282}$& $    -29451^{+   1412}_{-   2059}$&GE &*\\
BH 184    &   4.4  & $   41^{+ 14}_{- 17}$& $  121^{+  7}_{-  8}$& $  156^{+  7}_{-  7}$& $  4.7^{+0.2}_{-0.2}$& $ 1.69^{+0.15}_{-0.15}$& $0.47^{+0.04}_{-0.03}$& $  36^{+   2}_{-   2}$& $  58^{+   3}_{-   2}$& $    532^{+  39}_{-  37}$& $   -168541^{+   2788}_{-   2317}$&D &\\
NGC 6093   &   3.7  & $   33^{+  8}_{-  9}$& $   16^{+ 13}_{- 11}$& $   71^{+  8}_{-  5}$& $  4.8^{+0.0}_{-0.7}$& $ 0.06^{+0.38}_{-0.00}$& $0.98^{+0.00}_{-0.17}$& $  83^{+   4}_{-   6}$& $  44^{+   3}_{-   2}$& $     25^{+  29}_{-  17}$& $   -176940^{+   2815}_{-   1576}$&LE &\\
NGC 6121   &   6.3  & $  -52^{+  2}_{-  2}$& $    9^{+ 11}_{- 11}$& $   54^{+  4}_{-  2}$& $  6.5^{+0.1}_{-0.1}$& $ 0.12^{+0.10}_{-0.03}$& $0.96^{+0.01}_{-0.03}$& $  21^{+  76}_{-  31}$& $  68^{+   2}_{-   1}$& $     58^{+  69}_{-  70}$& $   -159029^{+   1144}_{-    910}$&LE &\\
NGC 6101   &  11.1  & $  -12^{+ 18}_{- 18}$& $ -312^{+  3}_{-  2}$& $  369^{+  3}_{-  4}$& $ 43.7^{+3.7}_{-3.6}$& $10.85^{+0.51}_{-0.51}$& $0.60^{+0.02}_{-0.01}$& $ 143^{+   1}_{-   1}$& $ 652^{+  61}_{-  60}$& $  -3217^{+ 163}_{- 152}$& $    -57030^{+   2609}_{-   2832}$&Seq &\\
NGC 6144   &   2.7  & $  -70^{+ 56}_{- 54}$& $ -197^{+ 34}_{- 14}$& $  213^{+  2}_{-  4}$& $  3.3^{+0.2}_{-0.1}$& $ 2.13^{+0.11}_{-0.20}$& $0.21^{+0.06}_{-0.02}$& $ 114^{+   4}_{-   4}$& $  40^{+  11}_{-  10}$& $   -240^{+  38}_{-  37}$& $   -172490^{+   2589}_{-   1903}$&Seq &\\
NGC 6139   &   3.5  & $   -1^{+ 11}_{- 19}$& $   74^{+  6}_{-  7}$& $  151^{+  7}_{-  6}$& $  3.6^{+0.3}_{-0.1}$& $ 1.01^{+0.20}_{-0.08}$& $0.56^{+0.03}_{-0.05}$& $  61^{+   3}_{-   3}$& $  50^{+   3}_{-   2}$& $    244^{+  31}_{-  25}$& $   -177317^{+   4002}_{-   1972}$&LE &\\
Terzan 3   &   2.5  & $  -60^{+ 38}_{- 36}$& $  205^{+  8}_{- 16}$& $  234^{+  6}_{-  7}$& $  3.2^{+0.3}_{-0.1}$& $ 2.20^{+0.22}_{-0.29}$& $0.18^{+0.08}_{-0.05}$& $  42^{+   4}_{-   4}$& $  44^{+   5}_{-   3}$& $    439^{+  48}_{-  49}$& $   -175727^{+   3297}_{-   2401}$&D &\\
NGC 6171   &   3.5  & $   -4^{+  3}_{-  2}$& $   77^{+ 10}_{-  8}$& $  101^{+  8}_{-  7}$& $  3.8^{+0.2}_{-0.2}$& $ 0.61^{+0.18}_{-0.14}$& $0.72^{+0.06}_{-0.07}$& $  52^{+   2}_{-   2}$& $  44^{+   3}_{-   2}$& $    189^{+  32}_{-  27}$& $   -179010^{+   2606}_{-   2445}$&B &\\
ESO 452-11   &   2.1  & $  -26^{+ 13}_{-  6}$& $  -14^{+  6}_{-  9}$& $  112^{+  3}_{-  8}$& $  2.9^{+0.1}_{-0.2}$& $ 0.06^{+0.01}_{-0.03}$& $0.96^{+0.02}_{-0.00}$& $ 101^{+   5}_{-   6}$& $  26^{+   2}_{-   1}$& $    -17^{+   7}_{-  11}$& $   -202137^{+   2802}_{-   2352}$&B &\\
NGC 6205   &   8.6  & $   21^{+  3}_{-  4}$& $  -26^{+  4}_{-  4}$& $   87^{+  3}_{-  3}$& $  8.6^{+0.2}_{-0.2}$& $ 1.01^{+0.11}_{-0.07}$& $0.79^{+0.01}_{-0.02}$& $ 105^{+   3}_{-   2}$& $ 100^{+   4}_{-   1}$& $   -185^{+  29}_{-  30}$& $   -134382^{+   1377}_{-   1006}$&GE &\\
NGC 6229   &  29.9  & $   30^{+  6}_{-  8}$& $   10^{+  2}_{-  3}$& $   58^{+  7}_{-  7}$& $ 31.0^{+1.0}_{-0.8}$& $ 0.64^{+0.29}_{-0.20}$& $0.96^{+0.01}_{-0.02}$& $  66^{+   8}_{-   6}$& $ 374^{+  14}_{-  10}$& $    231^{+  49}_{-  73}$& $    -74500^{+   1263}_{-   1091}$&GE &*\\
NGC 6218   &   4.8  & $   -9^{+  5}_{-  4}$& $  134^{+  5}_{-  6}$& $  157^{+  4}_{-  5}$& $  5.0^{+0.2}_{-0.2}$& $ 2.25^{+0.13}_{-0.16}$& $0.38^{+0.02}_{-0.03}$& $  37^{+   1}_{-   1}$& $  64^{+   4}_{-   1}$& $    578^{+  30}_{-  38}$& $   -158304^{+   1652}_{-   2303}$& D &\\
FSR 1735   &   4.3  & $  -80^{+ 10}_{- 12}$& $   20^{+ 13}_{-  8}$& $  174^{+ 10}_{-  5}$& $  4.8^{+0.4}_{-0.2}$& $ 0.43^{+0.19}_{-0.19}$& $0.84^{+0.07}_{-0.07}$& $  82^{+   3}_{-   5}$& $  58^{+   3}_{-   5}$& $     83^{+  53}_{-  35}$& $   -167696^{+   3898}_{-   4200}$&LE &*\\
NGC 6235   &   4.0  & $  159^{+  4}_{-  3}$& $  194^{+ 17}_{- 21}$& $  254^{+ 15}_{- 16}$& $  6.1^{+0.7}_{-0.8}$& $ 2.70^{+0.32}_{-0.40}$& $0.39^{+0.02}_{-0.02}$& $  53^{+   6}_{-   4}$& $  82^{+  11}_{-  10}$& $    561^{+ 110}_{- 141}$& $   -145991^{+   5819}_{-   7193}$&D &\\
NGC 6254   &   4.8  & $  -88^{+  3}_{-  3}$& $  134^{+  6}_{-  8}$& $  167^{+  4}_{-  5}$& $  5.2^{+0.2}_{-0.2}$& $ 2.13^{+0.13}_{-0.16}$& $0.42^{+0.03}_{-0.03}$& $  36^{+   2}_{-   1}$& $  78^{+   1}_{-   3}$& $    604^{+  26}_{-  41}$& $   -157521^{+   1403}_{-   2009}$&D &\\
NGC 6256   &   2.9  & $ -166^{+  9}_{-  8}$& $   34^{+ 23}_{- 24}$& $  194^{+  6}_{-  4}$& $  4.2^{+0.6}_{-0.4}$& $ 0.19^{+0.14}_{-0.13}$& $0.91^{+0.07}_{-0.07}$& $  76^{+  10}_{-   9}$& $  44^{+   5}_{-   2}$& $     94^{+  56}_{-  64}$& $   -182703^{+   6324}_{-   3401}$&LE &\\
Pal 15    &  38.2  & $  155^{+  8}_{-  8}$& $    4^{+ 12}_{-  9}$& $  162^{+  8}_{-  7}$& $ 54.5^{+4.5}_{-2.9}$& $ 1.32^{+1.37}_{-0.21}$& $0.95^{+0.01}_{-0.04}$& $  85^{+  15}_{-  18}$& $ 726^{+  72}_{-  45}$& $    119^{+ 414}_{- 329}$& $    -53297^{+   2569}_{-   1733}$&GE &*\\
NGC 6266   &   2.0  & $   41^{+  9}_{- 14}$& $  123^{+  9}_{-  9}$& $  146^{+  6}_{-  6}$& $  2.5^{+0.2}_{-0.4}$& $ 0.60^{+0.12}_{-0.11}$& $0.61^{+0.03}_{-0.06}$& $  32^{+   4}_{-   2}$& $  34^{+   3}_{-   5}$& $    217^{+  28}_{-  37}$& $   -205550^{+   4879}_{-   6500}$&B &\\
NGC 6273   &   1.6  & $  -98^{+ 88}_{- 97}$& $ -239^{+108}_{- 41}$& $  315^{+  5}_{-  5}$& $  3.8^{+0.5}_{-0.2}$& $ 1.00^{+0.21}_{-0.08}$& $0.59^{+0.02}_{-0.04}$& $ 109^{+  11}_{-  10}$& $  48^{+   6}_{-   2}$& $   -144^{+  77}_{-  96}$& $   -173042^{+   5314}_{-   2523}$&LE &\\
NGC 6284   &   7.3  & $   14^{+  2}_{-  2}$& $   -3^{+ 13}_{- 14}$& $  113^{+  6}_{-  6}$& $  7.5^{+0.7}_{-0.6}$& $ 0.72^{+0.16}_{-0.23}$& $0.82^{+0.05}_{-0.02}$& $  91^{+   8}_{-   7}$& $  90^{+   8}_{-   6}$& $    -19^{+  93}_{-  94}$& $   -142286^{+   5199}_{-   5012}$&GE &\\
NGC 6287   &   2.0  & $ -302^{+ 96}_{- 72}$& $  -64^{+ 13}_{- 18}$& $  319^{+  3}_{-  3}$& $  5.3^{+0.5}_{-0.3}$& $ 0.76^{+0.10}_{-0.06}$& $0.75^{+0.02}_{-0.03}$& $  95^{+   2}_{-   2}$& $  64^{+   7}_{-   3}$& $    -59^{+  26}_{-  28}$& $   -158899^{+   4804}_{-   2838}$&LE &\\
NGC 6293   &   1.8  & $ -152^{+  7}_{-  7}$& $  -80^{+ 12}_{- 17}$& $  232^{+  9}_{-  4}$& $  3.6^{+0.5}_{-0.3}$& $ 0.17^{+0.13}_{-0.04}$& $0.91^{+0.02}_{-0.05}$& $ 131^{+   7}_{-  11}$& $  38^{+   5}_{-   3}$& $    -93^{+  29}_{-  55}$& $   -191506^{+   7479}_{-   3379}$&B &\\
NGC 6304   &   2.5  & $   79^{+  5}_{-  6}$& $  190^{+  5}_{-  6}$& $  218^{+  4}_{-  5}$& $  3.3^{+0.3}_{-0.3}$& $ 1.79^{+0.20}_{-0.22}$& $0.29^{+0.02}_{-0.01}$& $  20^{+   1}_{-   1}$& $  52^{+   4}_{-   4}$& $    472^{+  56}_{-  55}$& $   -183230^{+   5448}_{-   5796}$&D &\\
NGC 6316   &   2.4  & $  102^{+  5}_{-  4}$& $   51^{+ 13}_{- 18}$& $  144^{+  8}_{-  7}$& $  3.0^{+0.4}_{-0.4}$& $ 0.40^{+0.14}_{-0.23}$& $0.76^{+0.13}_{-0.04}$& $  41^{+   8}_{-   4}$& $  36^{+   5}_{-   6}$& $    106^{+  45}_{-  50}$& $   -197198^{+   7404}_{-   9153}$&B &\\
NGC 6341   &   9.8  & $   50^{+  4}_{-  4}$& $   14^{+  4}_{-  5}$& $  108^{+  8}_{-  7}$& $ 10.6^{+0.2}_{-0.3}$& $ 0.44^{+0.16}_{-0.09}$& $0.92^{+0.02}_{-0.03}$& $  78^{+   4}_{-   4}$& $ 124^{+   3}_{-   4}$& $    121^{+  35}_{-  39}$& $   -125452^{+   1383}_{-   1633}$&GE &\\
NGC 6325   &   1.3  & $  -80^{+ 31}_{- 50}$& $ -178^{+131}_{- 59}$& $  209^{+ 23}_{- 23}$& $  1.3^{+0.5}_{-0.1}$& $ 1.06^{+0.13}_{-0.31}$& $0.11^{+0.22}_{-0.00}$& $ 114^{+  14}_{-  17}$& $  18^{+  11}_{-   3}$& $   -105^{+  78}_{-  86}$& $   -213001^{+   9843}_{-   4890}$&B &\\
NGC 6333   &   1.8  & $  -88^{+146}_{- 95}$& $  347^{+  8}_{- 42}$& $  364^{+  3}_{-  4}$& $  6.5^{+0.5}_{-0.3}$& $ 0.97^{+0.16}_{-0.14}$& $0.74^{+0.03}_{-0.03}$& $  59^{+   4}_{-   4}$& $  74^{+   6}_{-   2}$& $    327^{+  56}_{-  56}$& $   -151332^{+   3881}_{-   2911}$&LE &\\
NGC 6342   &   1.6  & $  -25^{+ 75}_{- 65}$& $  164^{+  8}_{- 42}$& $  169^{+  4}_{-  3}$& $  1.7^{+0.2}_{-0.1}$& $ 0.89^{+0.14}_{-0.17}$& $0.31^{+0.13}_{-0.08}$& $  64^{+   3}_{-   3}$& $  24^{+   7}_{-   2}$& $    117^{+  19}_{-  17}$& $   -206944^{+   3950}_{-   1304}$&B &\\
NGC 6356   &   7.2  & $   48^{+  6}_{-  5}$& $  105^{+ 16}_{- 19}$& $  156^{+ 11}_{-  9}$& $  7.9^{+0.7}_{-0.5}$& $ 2.37^{+0.57}_{-0.53}$& $0.54^{+0.08}_{-0.07}$& $  42^{+   5}_{-   3}$& $ 102^{+  11}_{-   6}$& $    701^{+ 136}_{- 140}$& $   -136904^{+   5232}_{-   4155}$&D &\\
NGC 6355   &   1.2  & $ -207^{+  7}_{-  9}$& $ -106^{+ 51}_{- 28}$& $  274^{+  5}_{-  6}$& $  2.2^{+0.8}_{-0.5}$& $ 0.64^{+0.09}_{-0.11}$& $0.55^{+0.12}_{-0.09}$& $ 106^{+   6}_{-   7}$& $  28^{+  10}_{-   5}$& $    -91^{+  48}_{-  56}$& $   -199376^{+  11298}_{-   9929}$&B &\\
NGC 6352   &   3.6  & $   43^{+ 11}_{- 10}$& $  226^{+  5}_{-  7}$& $  230^{+  4}_{-  5}$& $  4.1^{+0.3}_{-0.3}$& $ 3.17^{+0.14}_{-0.16}$& $0.13^{+0.03}_{-0.03}$& $  12^{+   1}_{-   1}$& $  68^{+   4}_{-   5}$& $    792^{+  50}_{-  48}$& $   -163943^{+   3332}_{-   3174}$&D &\\
IC 1257   &  17.6  & $  -48^{+  6}_{-  6}$& $  -37^{+  9}_{- 10}$& $   63^{+  6}_{-  3}$& $ 18.1^{+1.0}_{-0.8}$& $ 1.24^{+0.45}_{-0.32}$& $0.87^{+0.03}_{-0.04}$& $ 158^{+   2}_{-   4}$& $ 206^{+  11}_{-  10}$& $   -600^{+ 141}_{- 162}$& $    -99053^{+   2458}_{-   2292}$&GE &\\
\hline
  \end{tabular}
  \end{center}
  \end{scriptsize}
  \end{minipage}
  }

 \rotatebox{90}{
        \begin{minipage}{1.5\linewidth}
        \begin{scriptsize}
        \begin{center}
        \begin{tabular}{|l|r|r|r|r|r|r|r|r|r|r|r|r|r|}\hline
 Name &$d_{GC}$&$\Pi$ &$\Theta$ &$V_{tot}$ &apo   & peri      &ecc&incl.    &$T_r$&$L_Z$&$E$ &Galactic &\\
      & [kpc]  &[km/s]&[km/s]   &[km/s]    & [kpc]& [kpc]&   & $\theta$&[Gyr]&[kpc  &[km$^2$ & Sub-&\\
      &        &      &         &          &      &      &   & [deg]   &     & km/s]&/s$^2$&system&\\\hline
Terzan 2   &   1.0  & $ -120^{+ 30}_{- 17}$& $  -47^{+ 22}_{- 32}$& $  137^{+  3}_{-  2}$& $  1.2^{+0.4}_{-0.3}$& $ 0.09^{+0.04}_{-0.02}$& $0.86^{+0.07}_{-0.13}$& $ 160^{+   4}_{-  18}$& $  14^{+   5}_{-   4}$& $    -44^{+  13}_{-  13}$& $   -242209^{+  13782}_{-  14365}$&B &\\
NGC 6366   &   5.3  & $   94^{+  3}_{-  2}$& $  134^{+  3}_{-  4}$& $  175^{+  3}_{-  3}$& $  5.8^{+0.2}_{-0.2}$& $ 2.20^{+0.09}_{-0.11}$& $0.45^{+0.02}_{-0.01}$& $  32^{+   1}_{-   1}$& $  74^{+   3}_{-   1}$& $    697^{+  28}_{-  31}$& $   -153394^{+   1599}_{-   1663}$&D &\\
Terzan 4   &   1.2  & $   20^{+  9}_{- 16}$& $   67^{+ 11}_{-  9}$& $  118^{+  7}_{-  6}$& $  1.3^{+0.4}_{-0.3}$& $ 0.20^{+0.06}_{-0.03}$& $0.73^{+0.04}_{-0.09}$& $  54^{+   4}_{-   4}$& $  14^{+   4}_{-   4}$& $     82^{+  23}_{-  21}$& $   -235193^{+  11974}_{-  15449}$&B &\\
BH 229    &   0.5  & $    5^{+ 31}_{- 32}$& $  -54^{+ 25}_{-  5}$& $  293^{+ 12}_{- 14}$& $  0.8^{+0.7}_{-0.1}$& $ 0.28^{+0.12}_{-0.10}$& $0.49^{+0.23}_{-0.09}$& $ 100^{+   1}_{-   5}$& $  10^{+   6}_{-   1}$& $    -21^{+   8}_{-  10}$& $   -249619^{+  22805}_{-   1714}$&B &\\
FSR 1758   &   3.7  & $   63^{+ 26}_{- 27}$& $ -341^{+  6}_{-  4}$& $  401^{+  6}_{-  5}$& $ 13.8^{+1.9}_{-1.4}$& $ 3.66^{+0.40}_{-0.33}$& $0.58^{+0.02}_{-0.01}$& $ 147^{+   1}_{-   1}$& $ 174^{+  24}_{-  18}$& $  -1254^{+ 116}_{- 131}$& $   -108059^{+   6250}_{-   5485}$&Seq &\\
NGC 6362   &   5.2  & $   18^{+ 14}_{- 16}$& $  124^{+  9}_{-  7}$& $  160^{+  7}_{-  5}$& $  5.3^{+0.2}_{-0.1}$& $ 2.46^{+0.23}_{-0.17}$& $0.37^{+0.03}_{-0.03}$& $  45^{+   2}_{-   3}$& $  70^{+   5}_{-   1}$& $    583^{+  52}_{-  35}$& $   -153483^{+   2703}_{-   1632}$&D &\\
Liller 1   &   0.8  & $   97^{+ 20}_{- 26}$& $  -55^{+ 72}_{- 17}$& $  115^{+ 11}_{- 16}$& $  0.8^{+0.2}_{-0.0}$& $ 0.09^{+0.06}_{-0.07}$& $0.81^{+0.14}_{-0.07}$& $ 155^{+   8}_{-  89}$& $   8^{+   4}_{-   0}$& $    -41^{+  61}_{-  23}$& $   -262426^{+  12174}_{-   2514}$&B &\\
NGC 6380   &   3.1  & $  -59^{+  7}_{-  8}$& $  -29^{+ 11}_{- 10}$& $   67^{+ 10}_{-  8}$& $  3.4^{+0.4}_{-0.4}$& $ 0.16^{+0.21}_{-0.03}$& $0.91^{+0.01}_{-0.09}$& $ 168^{+   4}_{-   8}$& $  38^{+   6}_{-   3}$& $    -89^{+  39}_{-  47}$& $   -195003^{+   7157}_{-   5562}$&B &\\
Terzan 1   &   1.6  & $  -74^{+  2}_{-  4}$& $   65^{+  9}_{- 10}$& $   99^{+  8}_{-  6}$& $  1.8^{+0.3}_{-0.3}$& $ 0.21^{+0.07}_{-0.05}$& $0.78^{+0.03}_{-0.04}$& $   5^{+   3}_{-   1}$& $  22^{+   4}_{-   5}$& $    106^{+  26}_{-  28}$& $   -224589^{+   6942}_{-   9700}$&B &\\
Pismis 26   &   1.4  & $ -105^{+ 63}_{- 46}$& $  204^{+ 25}_{- 37}$& $  302^{+  7}_{-  7}$& $  3.0^{+0.5}_{-0.2}$& $ 0.92^{+0.43}_{-0.28}$& $0.54^{+0.14}_{-0.16}$& $  41^{+   4}_{-   2}$& $  42^{+  11}_{-   2}$& $    271^{+  62}_{-  50}$& $   -188443^{+   5667}_{-   2060}$&LE &\\
NGC 6388   &   3.0  & $  -65^{+ 12}_{- 17}$& $  -93^{+ 14}_{- 10}$& $  115^{+  7}_{-  5}$& $  3.5^{+0.2}_{-0.3}$& $ 0.62^{+0.22}_{-0.11}$& $0.70^{+0.03}_{-0.08}$& $ 148^{+   5}_{-   7}$& $  44^{+   4}_{-   2}$& $   -255^{+  49}_{-  45}$& $   -190244^{+   3550}_{-   2702}$&Seq &\\
NGC 6402   &   4.0  & $  -20^{+ 15}_{- 16}$& $   49^{+  5}_{-  7}$& $   56^{+  8}_{-  5}$& $  4.8^{+0.1}_{-0.4}$& $ 0.29^{+0.12}_{-0.02}$& $0.88^{+0.01}_{-0.04}$& $  46^{+   7}_{-   5}$& $  52^{+   3}_{-   3}$& $    159^{+  19}_{-  25}$& $   -176490^{+   2654}_{-   1937}$&LE &\\
NGC 6401   &   2.5  & $  -30^{+ 22}_{- 16}$& $ -255^{+  3}_{-  3}$& $  301^{+  5}_{-  4}$& $  4.5^{+0.6}_{-0.6}$& $ 2.37^{+0.46}_{-0.45}$& $0.31^{+0.04}_{-0.03}$& $ 144^{+   1}_{-   1}$& $  62^{+  12}_{-   9}$& $   -598^{+ 116}_{- 116}$& $   -162266^{+   8004}_{-   8586}$&Seq &\\
NGC 6397   &   6.3  & $   36^{+  4}_{-  5}$& $  127^{+  7}_{-  7}$& $  178^{+  6}_{-  6}$& $  6.5^{+0.1}_{-0.1}$& $ 2.76^{+0.20}_{-0.20}$& $0.41^{+0.03}_{-0.03}$& $  43^{+   2}_{-   2}$& $  84^{+   4}_{-   1}$& $    793^{+  41}_{-  42}$& $   -144589^{+   1350}_{-   1516}$&D &\\
Pal 6     &   2.5  & $ -191^{+  1}_{-  3}$& $   22^{+  8}_{- 11}$& $  245^{+  4}_{-  4}$& $  4.5^{+0.4}_{-0.6}$& $ 0.13^{+0.02}_{-0.06}$& $0.95^{+0.02}_{-0.02}$& $  83^{+   4}_{-   3}$& $  44^{+   4}_{-   4}$& $     55^{+  20}_{-  29}$& $   -179517^{+   4283}_{-   7078}$&LE &\\
NGC 6426   &  14.3  & $ -111^{+ 13}_{- 12}$& $   92^{+ 13}_{- 13}$& $  147^{+ 14}_{- 14}$& $ 16.6^{+0.9}_{-0.8}$& $ 3.17^{+0.67}_{-0.55}$& $0.68^{+0.04}_{-0.05}$& $  26^{+   3}_{-   3}$& $ 200^{+  17}_{-  12}$& $   1204^{+ 204}_{- 193}$& $   -100803^{+   3272}_{-   2975}$&H99 &\\
Djorg 1    &   1.2  & $ -255^{+161}_{- 65}$& $  311^{+ 59}_{- 34}$& $  402^{+  9}_{-  7}$& $  5.9^{+1.4}_{-1.7}$& $ 0.77^{+0.15}_{-0.07}$& $0.77^{+0.01}_{-0.07}$& $  24^{+   7}_{-   4}$& $  66^{+  15}_{-  16}$& $    346^{+  67}_{-  80}$& $   -162360^{+  11951}_{-  17508}$&GE &\\
Terzan 5   &   1.5  & $   76^{+  5}_{-  5}$& $   50^{+  8}_{- 11}$& $   96^{+  6}_{-  7}$& $  1.7^{+0.4}_{-0.3}$& $ 0.14^{+0.07}_{-0.03}$& $0.84^{+0.02}_{-0.04}$& $  40^{+   9}_{-   6}$& $  18^{+   6}_{-   4}$& $     75^{+  27}_{-  26}$& $   -228141^{+   9677}_{-   9909}$& B &\\
NGC 6440   &   1.3  & $   82^{+ 10}_{- 21}$& $  -41^{+ 30}_{- 29}$& $  100^{+  8}_{-  7}$& $  1.4^{+0.3}_{-0.0}$& $ 0.16^{+0.07}_{-0.09}$& $0.80^{+0.10}_{-0.06}$& $ 116^{+  18}_{-  20}$& $  14^{+   4}_{-   0}$& $    -46^{+  38}_{-  44}$& $   -232585^{+   6898}_{-    857}$& B &\\
NGC 6441   &   3.6  & $   16^{+  9}_{-  8}$& $   67^{+ 14}_{- 15}$& $   72^{+ 15}_{- 14}$& $  3.6^{+0.6}_{-0.4}$& $ 0.75^{+0.21}_{-0.21}$& $0.66^{+0.07}_{-0.06}$& $  22^{+   5}_{-   3}$& $  42^{+   7}_{-   4}$& $    230^{+  69}_{-  63}$& $   -186246^{+   7841}_{-   6993}$&LE &\\
Terzan 6   &   1.5  & $ -137^{+  3}_{-  3}$& $  -51^{+ 14}_{- 16}$& $  147^{+  5}_{-  3}$& $  1.9^{+0.4}_{-0.4}$& $ 0.15^{+0.07}_{-0.04}$& $0.85^{+0.03}_{-0.05}$& $ 170^{+   3}_{-   7}$& $  22^{+   7}_{-   4}$& $    -77^{+  26}_{-  28}$& $   -221047^{+  10098}_{-  11486}$&B &\\
NGC 6453   &   3.4  & $ -105^{+  6}_{-  5}$& $   39^{+ 13}_{- 14}$& $  197^{+  8}_{-  8}$& $  3.9^{+0.5}_{-0.7}$& $ 0.99^{+0.19}_{-0.14}$& $0.59^{+0.05}_{-0.07}$& $  78^{+   4}_{-   4}$& $  54^{+   6}_{-   8}$& $    132^{+  37}_{-  51}$& $   -172359^{+   6446}_{-   9017}$&LE &\\
NGC 6496   &   4.0  & $  -35^{+ 32}_{- 34}$& $  322^{+ 16}_{- 26}$& $  330^{+ 16}_{- 22}$& $  9.2^{+1.5}_{-1.4}$& $ 3.73^{+0.38}_{-0.35}$& $0.42^{+0.05}_{-0.06}$& $  32^{+   2}_{-   2}$& $ 120^{+  18}_{-  16}$& $   1118^{+ 129}_{- 144}$& $   -125956^{+   7076}_{-   7948}$&D &\\
Terzan 9   &   1.3  & $  -53^{+  9}_{-  6}$& $   15^{+ 15}_{-  9}$& $   82^{+  4}_{-  3}$& $  1.4^{+0.3}_{-0.4}$& $ 0.05^{+0.04}_{-0.02}$& $0.93^{+0.03}_{-0.08}$& $  78^{+   7}_{-  11}$& $  16^{+   3}_{-   5}$& $     19^{+  14}_{-  11}$& $   -235578^{+   9150}_{-  16419}$&B &\\
Djorg 2    &   2.0  & $  162^{+  5}_{-  3}$& $  159^{+  3}_{-  6}$& $  232^{+  2}_{-  3}$& $  3.2^{+0.3}_{-0.4}$& $ 0.91^{+0.14}_{-0.18}$& $0.56^{+0.03}_{-0.02}$& $  12^{+   1}_{-   1}$& $  42^{+   5}_{-   4}$& $    323^{+  43}_{-  58}$& $   -191864^{+   4864}_{-   7214}$&B &\\
NGC 6517   &   4.0  & $   47^{+  8}_{- 13}$& $   42^{+  9}_{-  6}$& $   73^{+  5}_{-  4}$& $  4.5^{+0.1}_{-0.3}$& $ 0.31^{+0.14}_{-0.04}$& $0.87^{+0.02}_{-0.06}$& $  52^{+   5}_{-   7}$& $  50^{+   3}_{-   2}$& $    161^{+  32}_{-  25}$& $   -179254^{+   2478}_{-   3018}$&LE &\\
Terzan 10   &   2.2  & $  229^{+  8}_{- 13}$& $   92^{+ 27}_{- 17}$& $  338^{+  8}_{-  9}$& $  5.8^{+0.8}_{-1.0}$& $ 0.68^{+0.14}_{-0.12}$& $0.79^{+0.02}_{-0.04}$& $  71^{+   3}_{-   5}$& $  74^{+  10}_{-  11}$& $    204^{+  24}_{-  25}$& $   -157214^{+   7702}_{-  10685}$&GE &\\
NGC 6522   &   0.8  & $   34^{+ 21}_{- 13}$& $   91^{+ 25}_{- 42}$& $  211^{+ 13}_{-  9}$& $  1.2^{+0.4}_{-0.2}$& $ 0.23^{+0.18}_{-0.13}$& $0.67^{+0.13}_{-0.10}$& $  63^{+  12}_{-   7}$& $  16^{+   9}_{-   5}$& $     58^{+  41}_{-  27}$& $   -238928^{+  19266}_{-  11639}$&B &\\
NGC 6535   &   4.0  & $   92^{+  7}_{-  8}$& $  -83^{+  8}_{-  7}$& $  131^{+  3}_{-  3}$& $  4.6^{+0.2}_{-0.2}$& $ 1.00^{+0.09}_{-0.09}$& $0.64^{+0.04}_{-0.02}$& $ 160^{+   1}_{-   1}$& $  56^{+   2}_{-   2}$& $   -320^{+  26}_{-  28}$& $   -173165^{+   2543}_{-   1988}$&Seq &\\
NGC 6528   &   0.7  & $ -196^{+177}_{- 58}$& $  111^{+ 64}_{- 31}$& $  229^{+  2}_{-  2}$& $  1.1^{+0.5}_{-0.5}$& $ 0.25^{+0.18}_{-0.06}$& $0.61^{+0.14}_{-0.39}$& $  71^{+   4}_{-   3}$& $  14^{+   5}_{-   4}$& $     50^{+  12}_{-  15}$& $   -241754^{+  12950}_{-  12284}$& B &\\
NGC 6539   &   3.1  & $   -2^{+ 15}_{- 18}$& $  118^{+  5}_{-  7}$& $  208^{+  9}_{-  6}$& $  3.4^{+0.2}_{-0.1}$& $ 1.86^{+0.23}_{-0.19}$& $0.29^{+0.04}_{-0.04}$& $  56^{+   2}_{-   2}$& $  56^{+   6}_{-   8}$& $    347^{+  26}_{-  28}$& $   -174408^{+   3585}_{-   2680}$&D &\\
NGC 6540   &   3.0  & $   13^{+  2}_{-  2}$& $  148^{+  5}_{-  6}$& $  159^{+  5}_{-  6}$& $  3.1^{+0.2}_{-0.3}$& $ 1.60^{+0.12}_{-0.18}$& $0.32^{+0.03}_{-0.03}$& $  21^{+   1}_{-   1}$& $  48^{+   2}_{-   4}$& $    450^{+  33}_{-  49}$& $   -187451^{+   3306}_{-   5125}$&D &\\
NGC 6544   &   5.3  & $    6^{+  2}_{-  2}$& $    6^{+ 12}_{-  7}$& $   91^{+  6}_{-  4}$& $  5.7^{+0.1}_{-0.3}$& $ 0.06^{+0.15}_{-0.03}$& $0.98^{+0.01}_{-0.05}$& $  86^{+   4}_{-   8}$& $  52^{+   4}_{-   1}$& $     30^{+  67}_{-  37}$& $   -166552^{+   1500}_{-   1861}$&GE &\\
NGC 6541   &   2.3  & $  124^{+ 23}_{- 43}$& $  192^{+ 22}_{- 16}$& $  253^{+  6}_{-  4}$& $  3.8^{+0.4}_{-0.4}$& $ 1.25^{+0.17}_{-0.10}$& $0.50^{+0.06}_{-0.08}$& $  40^{+   4}_{-   3}$& $  50^{+   4}_{-   4}$& $    333^{+  22}_{-  20}$& $   -175234^{+   3383}_{-   3409}$&LE &\\
ESO 280-06   &  13.8  & $   34^{+  7}_{-  6}$& $   27^{+ 16}_{- 13}$& $   86^{+ 10}_{-  6}$& $ 14.2^{+0.9}_{-0.9}$& $ 1.00^{+0.56}_{-0.36}$& $0.87^{+0.04}_{-0.06}$& $  67^{+  11}_{-  12}$& $ 164^{+  12}_{-  11}$& $    346^{+ 216}_{- 178}$& $   -110196^{+   3465}_{-   3298}$&GE &*\\
NGC 6553   &   2.4  & $   46^{+  8}_{-  6}$& $  246^{+  1}_{-  2}$& $  250^{+  1}_{-  1}$& $  3.3^{+0.2}_{-0.3}$& $ 2.27^{+0.27}_{-0.33}$& $0.19^{+0.03}_{-0.02}$& $   7^{+   1}_{-   1}$& $  52^{+   5}_{-   3}$& $    589^{+  67}_{-  80}$& $   -179761^{+   4706}_{-   6218}$& D &\\
NGC 6558   &   1.2  & $  187^{+  2}_{-  1}$& $   93^{+  9}_{-  9}$& $  210^{+  5}_{-  4}$& $  1.7^{+0.6}_{-0.4}$& $ 0.28^{+0.12}_{-0.06}$& $0.72^{+0.10}_{-0.14}$& $  63^{+   8}_{-   9}$& $  20^{+   6}_{-   4}$& $     88^{+  39}_{-  30}$& $   -217022^{+  10492}_{-   9416}$&B &\\
Pal 7     &   3.9  & $  -77^{+ 15}_{-  9}$& $  267^{+  7}_{-  6}$& $  279^{+  5}_{-  5}$& $  6.0^{+0.3}_{-0.4}$& $ 3.52^{+0.11}_{-0.17}$& $0.26^{+0.02}_{-0.02}$& $  11^{+   1}_{-   0}$& $  86^{+   4}_{-   4}$& $   1034^{+  40}_{-  59}$& $   -147446^{+   2348}_{-   3470}$&D &\\
\hline
  \end{tabular}
  \end{center}
  \end{scriptsize}
  \end{minipage}
  }

 \rotatebox{90}{
        \begin{minipage}{1.5\linewidth}
        \begin{scriptsize}
        \begin{center}
        \begin{tabular}{|l|r|r|r|r|r|r|r|r|r|r|r|r|r|r|}\hline
 Name &$d_{GC}$&$\Pi$ &$\Theta$ &$V_{tot}$ &apo   & peri &ecc&incl.    &$T_r$&$L_Z$&$E$ &Gal.& \\
      & [kpc]  &[km/s]&[km/s]   &[km/s]    & [kpc]& [kpc]&   & $\theta$&[Gyr]&[kpc  &[km$^2$ &Sub-&\\
      &        &      &         &          &      &      &   & [deg]   &     & km/s]&/s$^2$&syst&\\\hline
Terzan 12 &  3.6 & $ -97^{+  3}_{-  3}$& $  165^{+  6}_{-  7}$& $  213^{+  4}_{-  5}$& $  4.4^{+0.3}_{-0.3}$& $ 2.09^{+0.11}_{-0.14}$& $0.35^{+0.03}_{-0.01}$& $  29^{+   1}_{-   2}$& $  60^{+   4}_{-   2}$& $    599^{+  40}_{-  38}$& $   -169045^{+   3407}_{-   3215}$&D &\\
NGC 6569  &  2.8 & $ -40^{+  2}_{-  2}$& $  179^{+ 24}_{- 19}$& $  185^{+ 23}_{- 18}$& $  3.0^{+0.6}_{-0.4}$& $ 1.92^{+0.52}_{-0.44}$& $0.22^{+0.09}_{-0.06}$& $  26^{+   6}_{-   5}$& $  56^{+  10}_{-   8}$& $    451^{+ 133}_{- 101}$& $   -181599^{+   9717}_{-   7717}$&D &*\\
ESO 456-78&  2.0 & $  71^{+  9}_{-  7}$& $  198^{+  3}_{-  5}$& $  251^{+  5}_{-  6}$& $  2.9^{+0.3}_{-0.3}$& $ 1.43^{+0.31}_{-0.30}$& $0.34^{+0.06}_{-0.05}$& $  34^{+   1}_{-   2}$& $  52^{+   5}_{-   8}$& $    371^{+  69}_{-  63}$& $   -186882^{+   6566}_{-   7183}$&D &\\
NGC 6584  &  6.8 & $ 198^{+ 16}_{- 17}$& $  100^{+ 19}_{- 26}$& $  324^{+ 13}_{- 15}$& $ 18.0^{+2.1}_{-2.1}$& $ 1.68^{+0.51}_{-0.56}$& $0.83^{+0.05}_{-0.04}$& $  51^{+   6}_{-   4}$& $ 212^{+  27}_{-  27}$& $    565^{+ 141}_{- 175}$& $    -98062^{+   5083}_{-   6165}$&GE &\\
NGC 6624  &  1.2 & $ -29^{+ 44}_{- 20}$& $   59^{+  8}_{- 15}$& $  137^{+  7}_{-  4}$& $  1.6^{+0.2}_{-0.2}$& $ 0.13^{+0.14}_{-0.06}$& $0.85^{+0.06}_{-0.12}$& $  73^{+   2}_{-   4}$& $  22^{+   2}_{-   5}$& $     36^{+  17}_{-  13}$& $   -226289^{+   7818}_{-   3866}$&B &\\
NGC 6626  &  3.0 & $ -27^{+  3}_{-  3}$& $   57^{+  9}_{-  9}$& $  113^{+  6}_{-  6}$& $  3.1^{+0.3}_{-0.2}$& $ 0.45^{+0.09}_{-0.12}$& $0.75^{+0.07}_{-0.05}$& $  60^{+   4}_{-   4}$& $  42^{+   4}_{-   5}$& $    169^{+  30}_{-  29}$& $   -193005^{+   4073}_{-   3459}$&B &\\
NGC 6638  &  2.0 & $  68^{+  4}_{-  8}$& $   14^{+ 10}_{- 17}$& $   74^{+  4}_{-  6}$& $  2.4^{+0.4}_{-0.2}$& $ 0.05^{+0.05}_{-0.02}$& $0.96^{+0.02}_{-0.05}$& $  80^{+  13}_{-   7}$& $  20^{+   6}_{-   0}$& $     22^{+  13}_{-  27}$& $   -212090^{+   7283}_{-   2914}$&B &\\
NGC6637   &  1.6 & $  35^{+ 40}_{- 83}$& $   90^{+  4}_{- 42}$& $  126^{+  7}_{-  4}$& $  2.3^{+0.2}_{-0.3}$& $ 0.09^{+0.17}_{-0.06}$& $0.93^{+0.05}_{-0.15}$& $  77^{+   6}_{-   6}$& $  22^{+   2}_{-   1}$& $     40^{+  30}_{-  20}$& $   -212857^{+   4822}_{-   1774}$&B &\\
NGC 6642  &  1.7 & $ 112^{+  5}_{- 13}$& $   25^{+ 22}_{- 38}$& $  126^{+  5}_{-  8}$& $  2.2^{+0.2}_{-0.1}$& $ 0.08^{+0.07}_{-0.04}$& $0.93^{+0.03}_{-0.05}$& $  46^{+  65}_{-  19}$& $  24^{+   3}_{-   2}$& $     36^{+  40}_{-  54}$& $   -215381^{+   3709}_{-   1783}$&B &\\
NGC 6652  &  2.5 & $ -55^{+  4}_{-  1}$& $   30^{+ 11}_{- 16}$& $  184^{+  7}_{-  6}$& $  4.2^{+0.3}_{-0.2}$& $ 0.08^{+0.16}_{-0.07}$& $0.96^{+0.03}_{-0.06}$& $  75^{+   8}_{-   3}$& $  38^{+   6}_{-   2}$& $     46^{+  35}_{-  28}$& $   -183873^{+   6685}_{-   2374}$&B &\\
NGC 6656  &  5.2 & $ 176^{+  2}_{-  1}$& $  200^{+  1}_{-  1}$& $  302^{+  5}_{-  4}$& $  9.8^{+0.3}_{-0.3}$& $ 3.05^{+0.11}_{-0.09}$& $0.53^{+0.00}_{-0.01}$& $  33^{+   2}_{-   2}$& $ 126^{+   4}_{-   5}$& $   1041^{+  31}_{-  33}$& $   -125541^{+   1910}_{-   1690}$&D &\\
Pal 8     &  5.3 & $ -22^{+ 13}_{- 16}$& $  118^{+ 14}_{- 13}$& $  125^{+ 15}_{- 13}$& $  5.6^{+0.4}_{-0.6}$& $ 1.79^{+0.44}_{-0.26}$& $0.52^{+0.04}_{-0.09}$& $  24^{+   3}_{-   2}$& $  72^{+   6}_{-   6}$& $    601^{+ 103}_{-  89}$& $   -159674^{+   5561}_{-   4578}$&D &\\
NGC 6681  &  2.0 & $ 219^{+ 35}_{-120}$& $   57^{+113}_{- 50}$& $  287^{+  8}_{-  5}$& $  4.5^{+0.6}_{-0.4}$& $ 0.67^{+0.32}_{-0.20}$& $0.74^{+0.09}_{-0.12}$& $  84^{+   9}_{-   7}$& $  52^{+   5}_{-   2}$& $     37^{+  51}_{-  41}$& $   -168029^{+   4669}_{-   2227}$&LE &\\
NGC 6712  &  3.6 & $ 145^{+  3}_{-  2}$& $   25^{+ 13}_{- 10}$& $  208^{+  4}_{-  6}$& $  5.5^{+0.2}_{-0.4}$& $ 0.18^{+0.10}_{-0.08}$& $0.94^{+0.02}_{-0.04}$& $  79^{+   4}_{-   6}$& $  58^{+   2}_{-   5}$& $     92^{+  53}_{-  38}$& $   -168520^{+   3354}_{-   3335}$&GE &\\
NGC 6715  & 18.6 & $ 231^{+  5}_{-  4}$& $   51^{+ 14}_{- 13}$& $  313^{+  9}_{-  7}$& $ 53.9^{+10.6}_{-6.8}$& $14.66^{+1.02}_{-0.92}$& $0.57^{+0.04}_{-0.03}$& $  80^{+   3}_{-   3}$& $ 862^{+ 189}_{- 118}$& $    886^{+ 244}_{- 233}$& $    -49410^{+   4900}_{-   3976}$&Sgr &\\
NGC 6717  &  2.5 & $ -11^{+ 14}_{- 17}$& $  115^{+  5}_{-  9}$& $  118^{+  6}_{-  7}$& $  3.1^{+0.2}_{-0.5}$& $ 0.65^{+0.21}_{-0.06}$& $0.65^{+0.01}_{-0.10}$& $  32^{+   3}_{-   4}$& $  34^{+   5}_{-   1}$& $    249^{+  33}_{-  32}$& $   -195877^{+   4456}_{-   3703}$&B &\\
NGC 6723  &  2.6 & $ 101^{+ 12}_{-197}$& $ -179^{+334}_{- 17}$& $  209^{+  4}_{-  4}$& $  3.1^{+0.2}_{-0.2}$& $ 1.84^{+0.18}_{-0.12}$& $0.26^{+0.03}_{-0.03}$& $  90^{+   8}_{-   7}$& $  40^{+   4}_{-   1}$& $     -2^{+  66}_{-  75}$& $   -175214^{+   3304}_{-   2456}$&B &\\
NGC 6749  &  5.0 & $ -23^{+ 12}_{- 14}$& $  110^{+  8}_{-  8}$& $  112^{+  9}_{-  7}$& $  5.1^{+0.3}_{-0.2}$& $ 1.55^{+0.17}_{-0.15}$& $0.53^{+0.04}_{-0.03}$& $   3^{+   0}_{-   0}$& $  62^{+   4}_{-   1}$& $    555^{+  45}_{-  41}$& $   -167074^{+   3062}_{-   2247}$& D &\\
NGC 6752  &  5.5 & $ -24^{+  3}_{-  3}$& $  179^{+  5}_{-  3}$& $  190^{+  5}_{-  3}$& $  5.7^{+0.2}_{-0.2}$& $ 3.57^{+0.21}_{-0.15}$& $0.23^{+0.01}_{-0.02}$& $  24^{+   1}_{-   1}$& $  82^{+   4}_{-   2}$& $    932^{+  47}_{-  38}$& $   -147118^{+   1997}_{-   1841}$& D &\\
NGC 6760  &  5.0 & $  91^{+ 10}_{- 10}$& $  147^{+  6}_{-  6}$& $  173^{+  5}_{-  5}$& $  5.6^{+0.2}_{-0.2}$& $ 2.17^{+0.19}_{-0.18}$& $0.44^{+0.03}_{-0.03}$& $   6^{+   0}_{-   0}$& $  72^{+   3}_{-   3}$& $    725^{+  52}_{-  50}$& $   -158822^{+   2626}_{-   2401}$& D &\\
NGC 6779  &  9.3 & $ 154^{+  1}_{-  1}$& $  -14^{+  6}_{-  4}$& $  184^{+  3}_{-  3}$& $ 12.5^{+0.5}_{-0.7}$& $ 0.25^{+0.15}_{-0.06}$& $0.96^{+0.01}_{-0.02}$& $ 100^{+   3}_{-   4}$& $ 134^{+   6}_{-   8}$& $   -129^{+  57}_{-  40}$& $   -119887^{+   2614}_{-   3009}$&GE &\\
Terzan 7  & 15.3 & $ 259^{+  6}_{-  6}$& $   32^{+ 14}_{- 15}$& $  319^{+  9}_{-  8}$& $ 42.3^{+6.5}_{-4.7}$& $12.73^{+0.75}_{-0.68}$& $0.54^{+0.03}_{-0.03}$& $  85^{+   3}_{-   2}$& $ 656^{+ 109}_{-  75}$& $    422^{+ 180}_{- 200}$& $    -56985^{+   4381}_{-   3760}$&Sgr& \\
Pal 10    &  6.6 & $ -64^{+ 10}_{- 10}$& $  187^{+  8}_{-  8}$& $  198^{+  9}_{-  8}$& $  7.2^{+0.3}_{-0.2}$& $ 4.00^{+0.29}_{-0.25}$& $0.28^{+0.03}_{-0.02}$& $   7^{+   0}_{-   0}$& $ 100^{+   5}_{-   3}$& $   1240^{+  69}_{-  60}$& $   -137907^{+   2591}_{-   2050}$&D &\\
Arp 2     & 21.2 & $ 242^{+  6}_{-  8}$& $   61^{+ 18}_{- 11}$& $  306^{+  8}_{-  7}$& $ 61.3^{+9.4}_{-7.0}$& $17.54^{+0.89}_{-0.77}$& $0.55^{+0.04}_{-0.03}$& $  79^{+   2}_{-   3}$& $1026^{+ 169}_{- 125}$& $   1128^{+ 327}_{- 213}$& $    -45162^{+   3483}_{-   3081}$&Sgr &\\
NGC 6809  &  4.1 & $-199^{+  3}_{-  4}$& $   75^{+ 11}_{- 13}$& $  220^{+  3}_{-  2}$& $  5.7^{+0.3}_{-0.3}$& $ 1.18^{+0.15}_{-0.14}$& $0.66^{+0.04}_{-0.04}$& $  68^{+   4}_{-   3}$& $  76^{+   3}_{-   3}$& $    262^{+  41}_{-  46}$& $   -154418^{+   2721}_{-   2294}$&LE &\\
Terzan 8  & 19.1 & $ 269^{+  7}_{-  6}$& $   41^{+ 12}_{- 17}$& $  317^{+  9}_{-  7}$& $ 60.0^{+9.7}_{-6.9}$& $16.12^{+0.88}_{-0.76}$& $0.58^{+0.03}_{-0.04}$& $  83^{+   3}_{-   2}$& $ 982^{+ 174}_{- 121}$& $    646^{+ 171}_{- 267}$& $    -46155^{+   3726}_{-   3177}$&Sgr &\\
Pal 11    &  8.1 & $ -17^{+ 15}_{- 19}$& $  138^{+  9}_{- 12}$& $  140^{+ 11}_{- 12}$& $  8.2^{+0.4}_{-0.3}$& $ 3.47^{+0.37}_{-0.45}$& $0.40^{+0.06}_{-0.04}$& $  27^{+   2}_{-   2}$& $ 110^{+   7}_{-   5}$& $   1012^{+  81}_{- 107}$& $   -132025^{+   2555}_{-   2704}$&D &\\
NGC 6838  &  7.0 & $  38^{+  5}_{-  5}$& $  204^{+  3}_{-  2}$& $  211^{+  2}_{-  2}$& $  7.3^{+0.2}_{-0.1}$& $ 5.00^{+0.13}_{-0.11}$& $0.18^{+0.02}_{-0.01}$& $  12^{+   0}_{-   1}$& $ 110^{+   3}_{-   1}$& $   1422^{+  28}_{-  25}$& $   -132364^{+   1077}_{-    941}$&D &\\
NGC 6864  & 14.6 & $ -98^{+  8}_{-  7}$& $   21^{+ 11}_{- 13}$& $  112^{+  8}_{-  8}$& $ 16.5^{+0.8}_{-0.7}$& $ 0.59^{+0.47}_{-0.23}$& $0.93^{+0.03}_{-0.05}$& $  56^{+  20}_{-  13}$& $ 186^{+  10}_{-   9}$& $    238^{+ 136}_{- 147}$& $   -103497^{+   2440}_{-   2223}$&GE & \\
NGC 6934  & 12.7 & $-290^{+ 14}_{- 13}$& $  105^{+ 20}_{- 22}$& $  332^{+ 12}_{- 12}$& $ 41.4^{+5.4}_{-5.2}$& $ 2.57^{+0.74}_{-0.50}$& $0.88^{+0.02}_{-0.03}$& $  23^{+   3}_{-   1}$& $ 528^{+  83}_{-  76}$& $   1224^{+ 228}_{- 267}$& $    -62807^{+   4391}_{-   4984}$&GE &\\
NGC 6981  & 12.8 & $-158^{+ 11}_{-  9}$& $    2^{+ 15}_{- 11}$& $  232^{+  7}_{-  7}$& $ 22.5^{+1.2}_{-0.8}$& $ 0.38^{+0.10}_{-0.23}$& $0.97^{+0.02}_{-0.01}$& $  79^{+  29}_{-  27}$& $ 254^{+  17}_{-   9}$& $     15^{+ 141}_{- 103}$& $    -89148^{+   2531}_{-   1655}$&GE &*\\
NGC 7006  & 38.5 & $-144^{+  5}_{-  4}$& $  -20^{+  7}_{-  7}$& $  167^{+  4}_{-  6}$& $ 56.5^{+3.1}_{-3.2}$& $ 1.72^{+0.67}_{-0.48}$& $0.94^{+0.02}_{-0.02}$& $ 125^{+  11}_{-  10}$& $ 758^{+  52}_{-  54}$& $   -725^{+ 241}_{- 249}$& $    -52150^{+   1759}_{-   1956}$&GE &*\\
NGC 7078  & 10.6 & $   6^{+  5}_{-  7}$& $  119^{+  6}_{-  6}$& $  123^{+  6}_{-  6}$& $ 10.6^{+0.4}_{-0.3}$& $ 3.60^{+0.29}_{-0.26}$& $0.49^{+0.03}_{-0.02}$& $  28^{+   2}_{-   1}$& $ 142^{+   4}_{-   6}$& $   1128^{+  70}_{-  69}$& $   -119827^{+   1919}_{-   1789}$&D &\\
NGC 7089  & 10.4 & $ 168^{+  5}_{-  5}$& $  -15^{+  8}_{-  5}$& $  240^{+  9}_{-  6}$& $ 18.5^{+1.2}_{-0.9}$& $ 0.55^{+0.04}_{-0.24}$& $0.94^{+0.03}_{-0.00}$& $ 116^{+   9}_{-  13}$& $ 208^{+  16}_{-  11}$& $   -120^{+  61}_{-  41}$& $    -98441^{+   3033}_{-   2478}$&GE &\\
NGC 7099  &  7.2 & $ -34^{+  9}_{-  8}$& $  -55^{+ 12}_{- 10}$& $  127^{+  7}_{-  6}$& $  8.2^{+0.4}_{-0.2}$& $ 0.98^{+0.24}_{-0.20}$& $0.79^{+0.04}_{-0.05}$& $ 119^{+   3}_{-   5}$& $  94^{+   5}_{-   3}$& $   -233^{+  54}_{-  45}$& $   -137334^{+   2437}_{-   1730}$&GE &\\
Pal 12    & 15.7 & $ 143^{+ 20}_{- 28}$& $  304^{+ 17}_{- 16}$& $  354^{+ 14}_{- 14}$& $ 69.9^{+18.2}_{-14.3}$& $15.49^{+0.72}_{-0.55}$& $0.64^{+0.05}_{-0.06}$& $  67^{+   1}_{-   2}$& $1140^{+ 297}_{- 313}$& $   2142^{+ 144}_{- 145}$& $    -42546^{+   5008}_{-   5202}$& Sgr &\\
Pal 13    & 27.0 & $ 264^{+  5}_{-  5}$& $  -92^{+ 10}_{- 11}$& $  293^{+  7}_{-  6}$& $ 90.9^{+8.3}_{-7.2}$& $ 8.31^{+0.93}_{-0.64}$& $0.83^{+0.01}_{-0.01}$& $ 120^{+   4}_{-   4}$& $1418^{+ 164}_{- 138}$& $  -1876^{+ 238}_{- 258}$& $    -37614^{+   2142}_{-   2097}$&GE &\\
NGC 7492  & 25.4 & $ -91^{+ 18}_{- 13}$& $   -6^{+  6}_{-  5}$& $  112^{+ 11}_{- 14}$& $ 28.3^{+3.3}_{-2.6}$& $ 3.34^{+1.20}_{-1.05}$& $0.79^{+0.06}_{-0.06}$& $  92^{+   2}_{-   3}$& $ 354^{+  29}_{-  22}$& $    -57^{+  59}_{-  45}$& $    -76660^{+   2636}_{-   2436}$& GE& *\\
\hline
   \end{tabular}
   \end{center}
   \end{scriptsize}
   \end{minipage}
  }
\end{document}